An Adaptive Real-Time Forecasting Framework for Cryogenic Fluid Management in Space Systems

Qiyun Cheng, Huihua Yang, and Wei Ji[*]

Department of Mechanical, Aerospace, and Nuclear Engineering, Rensselaer Polytechnic Institute, 110 8th Street, Troy, NY 12018

## Abstract

Accurate real-time forecasting of cryogenic tank behavior is essential for the safe and efficient operation of propulsion and storage systems in future deep-space missions. While cryogenic fluid management (CFM) systems increasingly require autonomous capabilities, conventional simulation methods remain hindered by high computational cost, model imperfections, and sensitivity to unanticipated boundary condition changes. To address these limitations, this study proposes an Adaptive Real-Time Forecasting Framework for Cryogenic Propellant Management in Space Systems, featuring a lightweight, non-intrusive method named ARCTIC (Adaptive Real-time Cryogenic Tank Inference and Correction). ARCTIC integrates real-time sensor data with precomputed nodal simulations through a data-driven correction layer that dynamically refines forecast accuracy without modifying the underlying model. Two updating mechanisms, auto-calibration and observation and correction, enable continuous adaptation to evolving system states and transient disturbances. The method is first assessed through synthetic scenarios representing self-pressurization, sloshing, and periodic operations, then validated using experimental data from NASA's Multipurpose Hydrogen Test Bed and K-Site facilities. Results demonstrate that ARCTIC significantly improves forecast accuracy under model imperfections, data noise, and boundary fluctuations, offering a robust real-time forecasting capability to support autonomous CFM operations. The framework's compatibility with existing simulation tools and its low computational overhead make it especially suited for onboard implementation in space systems requiring predictive autonomy.

**Nomenclature**

**Latin Letters**

| | |
|---|---|
| $A$ | Area [$\mathrm{m}^2$] |
| $b$ | Maximum sloshing amplitude [m] |
| $C$ | Constant [-] |
| $F$ | Friction coefficient [-] |
| $g$ | Gravitational acceleration [$\mathrm{m} \cdot \mathrm{s}^{-2}$] |
| $H$ | Specific enthalpy [$\mathrm{J} \cdot \mathrm{kg}^{-1}$] |
| $h$ | Heat transfer coefficient [$\mathrm{W} \cdot \mathrm{m}^{-2} \cdot \mathrm{K}$] |
| $h_{fg}$ | Latent heat of vaporization [$\mathrm{J} \cdot \mathrm{kg}^{-1}$] |
| $K$ | K-factor for losses [-] |
| $k$ | Thermal conductivity [$\mathrm{W} \cdot \mathrm{m}^{-1} \cdot \mathrm{K}$] |
| $L$ | Length |
| $m$ | Mass [kg] |
| $\dot{m}$ | Mass flow rate [$\mathrm{kg} \cdot \mathrm{s}^{-1}$] |
| $n$ | Flow direction [-] |
| $\bar{v}$ | Average velocity [$\mathrm{m} \cdot \mathrm{s}^{-1}$] |
| $x$ | Local elevation [m] |

**Greek Letters**

| | |
|---|---|
| $\beta$ | Thermal expansion coefficient [$\mathrm{K}^{-1}$] |
| $\delta$ | Boundary layer thickness [m] |
| $\rho$ | Density [$\mathrm{kg} \cdot \mathrm{m}^{-3}$] |
| $\nu$ | Kinematic viscosity [$\mathrm{m}^2 \cdot \mathrm{s}^{-1}$] |
| $\eta$ | Frequency ratio [-] |

**Non-dimensional Numbers**

| | |
|---|---|
| Gr | Grashof number [-] |
| Nu | Nusselt number [-] |
| Pr | Prandtl number [-] |
| Ra | Rayleigh number [-] |

[*] Corresponding author: jiw2@rpi.edu

| | | | |
|---|---|---|---|
| $p$ | Pressure [Pa] | Re | Reynolds number [-] |
| $R$ | Radius [m] | $(\mathrm{Re}_s)_c$ | Critical Reynolds number [-] |
| $T$ | Temperature [K] | | |
| $T_{sat}$ | Saturation temperature [K] | **Subscripts** | |
| $t$ | Time [s] | $i,j$ | Lump/tube index |
| $\Delta t_D$ | Data collection window [s] | $l$ | Liquid |
| $\Delta t_{AC}$ | Auto Calibration period [s] | $s$ | Vapor |
| $U$ | Internal energy [J] | $v$ | Sloshing |

## 1. Introduction

Cryogenic fluids are essential in space missions, serving as both propellants and life-support resources for launch vehicles and deep-space systems. Liquid hydrogen ($LH_2$), for example, is widely used as a high-performance propellant due to its high specific impulse and favorable thrust-to-weight ratio [1]. However, the cryogenic nature of these fluids presents significant challenges for long-duration in-space storage and handling, especially in autonomous missions where ground intervention is limited or unavailable [2,3].

Cryogenic fuels must be maintained at a very low temperature (usually below $-150$℃) to remain in the liquid phase, making their storage highly sensitive to environmental heat leakage and dynamic disturbances. During the long-term storage, even with advanced insulation, cryogenic storage tanks aboard space vehicles continuously absorb heat from background radiation and structural conduction, leading to gradual self-pressurization and thermal stratification over time [4]. These effects accumulate and necessitate periodic depressurization operations to maintain tank safety. In addition to heat leakage, sloshing presents another major challenge. Vehicle maneuvers and mechanical oscillations induce sloshing of the cryogenic liquid [5-7], which can significantly enhance interfacial heat and mass transfer. During sloshing, even small-amplitude motions can cause rapid condensation or vaporization when cold liquid contacts hot vapor or heated tank walls, producing sudden pressure collapse or spikes and reducing thermal-pressure stability. To maintain safe operating conditions for both the storage tank and the associated propulsion systems, pressure regulation through vapor injection or venting is often required. While these control actions may be mechanically straightforward, determining the appropriate timing and magnitude of such operations is complicated by the underlying physics. Boil-off losses, nonlinear phase-change dynamics, and unpredictable two-phase flow behavior, particularly under microgravity, can lead to rapid and difficult-to-predict changes in pressure and temperature. These phenomena challenge the control system’s ability to respond accurately and efficiently, underscoring the need for a reliable and adaptive cryogenic fluid management (CFM) system capable of handling a wide range of operating conditions [3,8].

A CFM system is responsible for maintaining thermal and pressure stability in cryogenic storage tanks by managing boil-off, pressure rise, stratification, and sloshing-induced disturbances. Such systems typically employ venting, mixing, and other active control mechanisms to regulate tank pressure and fluid stratification. In current practice, these operations rely predominantly on threshold-based control mechanisms or ground-commanded interventions. Threshold-based mechanisms trigger control actions when parameters such as tank pressure, temperature, or liquid level are exceeded predefined limits [9]. Although straightforward to implement, recent studies have shown that purely threshold-based control can introduce substantial delays, unnecessary vent cycles, and excessive propellant loss [14]. Thermodynamic vent system (TVS) experiments [14] demonstrate that the inherent mismatch in the response times of pressure and temperature in cryogenic tanks can cause different threshold strategies to produce several-fold differences in propellant loss. Moreover, due to the complex multiphase behavior of cryogenic propellants, recent operational analyses further emphasize that sloshing-induced transients and microgravity convection can mask early indicators of unsafe conditions, meaning that threshold-driven controllers may either over-react to benign disturbances or fail to respond promptly to emerging risks [15]. Ground-based interventions, on the other hand, become impractical for future deep-space missions due to significant communication delays, such as the 5 to 20 minutes one-way delay associated with Mars exploration [12]. Therefore, as space exploration moves toward more complex and long-duration missions, the integration of proactive, autonomous control systems is imperative to minimize human intervention. These systems can help achieve greater fuel efficiency, enhanced mission reliability, and increase resilience against unforeseen in-space conditions [13]. Recent NASA roadmaps and international CFM studies uniformly identify autonomous, predictive control as a key enabling capability for in-space cryogenic logistics, depots, and exploration missions [8, 16, 17].

To enable autonomous control of CFM systems, accurate and fast simulations are essential for predicting the thermodynamic evolution and enabling the control system to anticipate future system states and take action ahead of time. Lumped parameter nodal (system) codes [18] are widely used due to their computational efficiency and flexibility, particularly for long-duration simulations. However, their underlying 0D/1D formulation and coarse spatial resolution limit their ability to capture inherently three-dimensional (3D) phenomena such as natural convection, stratification, and jet-induced mixing [19, 20]. Significant efforts have been made to enhance the fidelity of nodal simulations through incorporating empirical or semi-empirical correlations [11, 21-23], multi-code coupling strategies [24,25,26], and data assimilation techniques [27]. Empirical correlations are commonly embedded within nodal models to model sub-grid physics, such as using convective heat transfer correlations to represent liquid-vapor interfacial heat and mass transfer. Several research, such as [11, 21-23], demonstrated that appropriately formulated correlations can enhance the accuracy of nodal predictions for specific operating scenarios, including no-vent fill, self-pressurization, and upper-stage tank operations. While these correlations improve predictive accuracy, these correlations are generally developed and validated under particular flow regimes or gravity conditions, and their applicability outside those regimes, especially in microgravity, where experimental data remain limited [2], is not yet fully established. Integrated analysis by coupling the nodal codes with high fidelity computational fluid dynamic (CFD) codes is another approach to tackle the complex 3D effects [24,25,26]. However, because CFD substantially increases computational cost relative to stand-alone nodal models, its suitability for real-time or on-board implementation remains a challenge, particularly for long prediction horizons. Data assimilation strategies have also been explored. Marques et al. used artificial neural networks (ANNs) to estimate heat transfer coefficients from real-time data [27]. Although their method showed promise on synthetic datasets by successfully learning predefined empirical correlations, it lacks experimental validations. More importantly, its reliance on ANN-based closures trained on-the-fly, while simultaneously simulating the system using nodal codes, imposing significant computational demands. This limits its applications for computational resource-constrained platforms such as spaceflight systems, where real-time prediction is essential but difficult to achieve with such a computationally intensive framework. In previous studies, empirical correlations, high fidelity simulations, and ANNs are typically used to calibrate individual physical parameters, such as interfacial heat transfer coefficients. However, this parameter-centric approach is often inadequate for CFM systems, especially under microgravity conditions, where unmodeled physical phenomena can lead to significant discrepancies, as revealed by in-space experiments [3]. Furthermore, the error of the numerical simulations in real-world applications can also arise from uncertainties and fluctuations of initial and boundary conditions, as well as geometric simplifications. Artificial neural networks have also been used to directly predict the thermodynamic state of cryogenic tanks as surrogates for physics-based numerical solvers [43,44]. Such models can provide fast and accurate predictions within the range of the available training data. However, due to the intrinsic interpolation nature of neural networks, their reliability in extrapolating to unseen operational scenarios remains an open research challenge. These challenges highlight the need for more adaptive methods that go beyond parameter calibration to account for the full range of physical variability, ultimately enabling CFM systems to confidently utilize model forecasts for proactive adjustment of control actions.

To address these challenges, we propose ARCTIC (Adaptive Real-time Cryogenic Tank Inference and Correction), a real-time forecasting framework designed to enhance the accuracy of nodal simulations by assimilating real-time sensor data in a non-intrusive and adaptive manner. In ARCTIC, mission-specific nodal simulations are precomputed offline, generating offline predictions. During the flight, a digital monitoring agent continuously collects sensor data, detects deviations between predictions and measurements, and updates a set of lightweight, data-driven correction functions. These functions directly map key tank variables, such as pressure and temperature, from the offline nodal outputs to observed sensor data, implicitly accounting for model imperfections, boundary fluctuations, and unmodeled cryogenic phenomena. Since these data-driven corrections are applied directly to the offline predictions as an additional layer for data-assimilation, the original nodal model and codes remain unchanged, and no intrusive modification of the underlying model is required. Two updating mechanisms are integrated into ARCTIC: an auto-calibration routine that periodically refines correlations under steady conditions, and an observation-and-correction strategy that re-trains mappings when deviations exceed a predefined threshold. Forecasts are produced over a tunable prediction window based on the response time requirements of the control system, enabling proactive adjustments ahead of critical events. By providing accurate real-time forecasts without modifying the nodal mode, ARCTIC supports predictive autonomous control systems and is especially suitable for flight software environments with limited computational resources.

The main contributions of this work are as follows:

1. We develop ARCTIC, an adaptive, non-intrusive forecasting framework that integrates offline nodal simulations with real-time measurement data to enable fast and accurate thermodynamic state predictions in cryogenic propellant tanks.
2. The framework incorporates a dual updating mechanism that combines periodic auto-calibration with event-triggered observation and correction, allowing the model to adapt dynamically to both gradual system changes and fast transients such as maneuver-induced sloshing.
3. A lightweight, correlation-based mapping strategy is implemented to compensate for model imperfections, uncertain boundary conditions, and unmodeled cryogenic behaviors without parameter tuning or numerical model modification.
4. ARCTIC is thoroughly evaluated using synthetic test scenarios and validated with experimental data from NASA's Multipurpose Hydrogen Test Bed (MHTB) and K-Site facilities, demonstrating robust performance and real-time applicability for supporting autonomous CFM systems.

The remainder of this paper is structured as follows: Section 2 details the proposed ARCTIC framework. Section 3 presents the nodal models and empirical correlations used in this work. Section 4 evaluates ARCTIC's capability to capture multiple unmodeled cryogenic behaviors and boundary fluctuations using synthetic data under different scenarios, including self-pressurization, holding, periodic venting and mixing, and sloshing. The impact of data noise is also discussed. Section 5 validates the approach using experimental data. The self-pressurization experiment of Multipurpose Hydrogen Test Bed (MHTB) [28,29] is first used to validate the framework, followed by the self-pressurization experiment of the large cryogenic tank in K-Site facility [30,31], as well as the sloshing experiment of the small tank in K-Site facility [32]. Section 6 demonstrates the framework's applicability in a control-oriented case study. Section 7 summarizes the findings and discusses the potential and future directions of ARCTIC for supporting future autonomous space systems.

## 2. The ARCTIC Framework: Adaptive Real-Time Forecasting for Cryogenic Tanks

In a typical space mission, the geometry of the cryogenic propellant tank is predefined, and the mission profile is well-planned prior to launch. Consequently, a nodal model of the cryogenic storage tank can be constructed using initial and boundary conditions based on the mission plan to simulate the entire storage stage of the mission, named as "offline predictions". After the space shuttle is launched and the cryogenic tank enters the long-term storage stage, a real-time monitoring agent starts to continuously collect sensor data from the cryogenic propellant tank ("environment"). Discrepancies between offline predictions and real-time sensor data are usually observed due to model imperfections, deviations from planned boundary conditions, and unmodeled cryogenic behaviors [3]. To address these differences, the monitoring agent can keep calibrating and correcting the predictions based on the real-time sensor data. The resulting data-informed predictions provide a highly accurate representation of the thermodynamic state of the environment, making them suitable for integration with other onboard systems, including autonomous control systems. Figure 1 illustrates the overall structure of the ARCTIC framework. Due to its non-intrusive design, the ARCTIC can be seamlessly integrated with any numerical simulation code without requiring modifications.

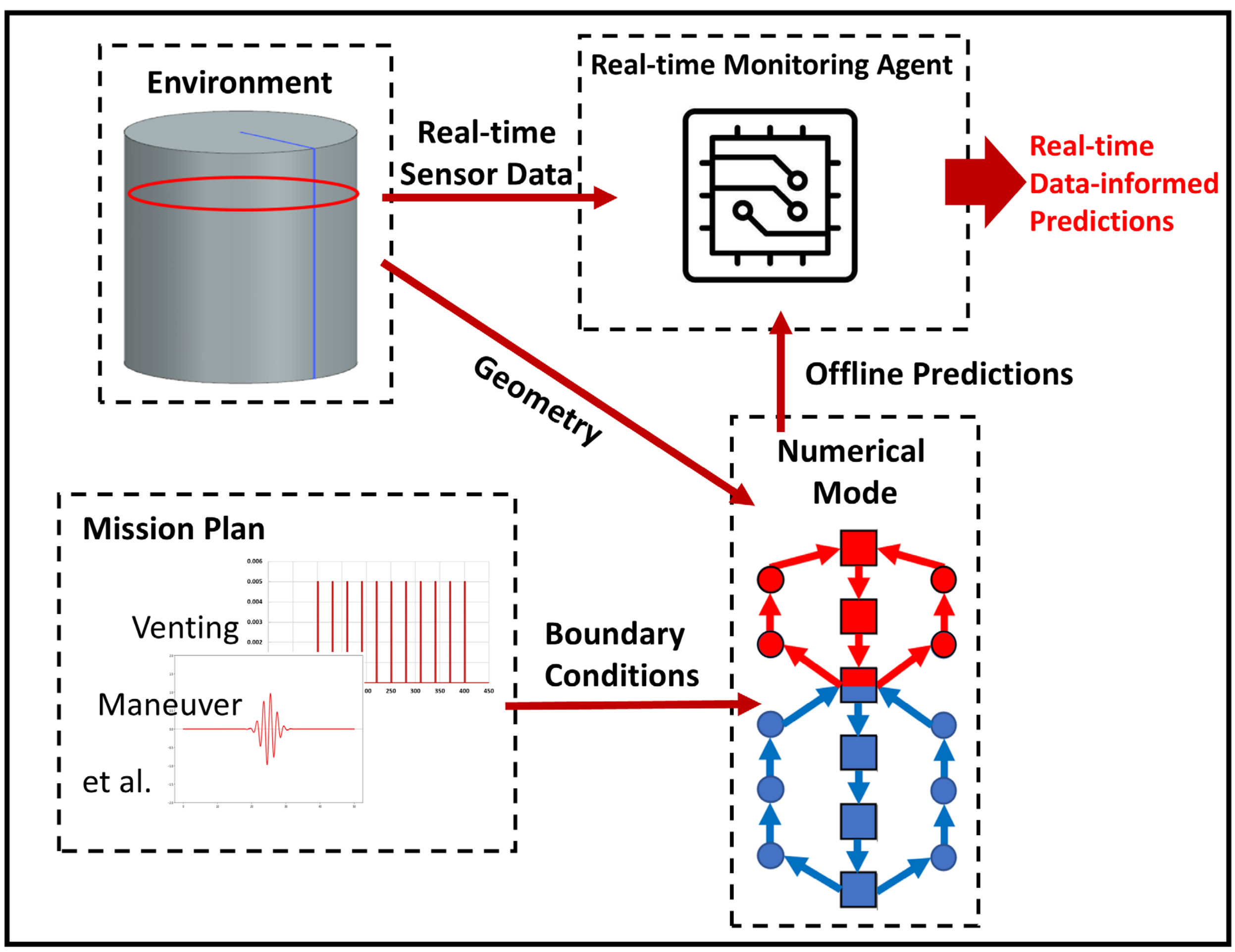

Figure 1. Overview of the ARCTIC framework and its interaction with the cryogenic tank system. A set of offline nodal simulations, constructed using the mission plan (operations), tank geometry, and expected boundary conditions, are generated prior to the mission to provide baseline thermodynamic evolution curves. During operation, the cryogenic tank ("Environment") supplies real-time sensor measurements of key thermodynamic quantities. These measurements stream into the Monitoring Agent, which evaluates discrepancies between offline predictions and sensor data and continuously corrects the baseline predictions through lightweight, data-driven mappings. The resulting corrected outputs ("Data-informed Predictions") deliver accurate real-time state estimates and future forecasts, enabling downstream systems such as autonomous thermal and pressure control to make proactive decisions.

To ensure clarity, the key operations of the ARCTIC framework used in this study are defined as follows:

- **Data-driven Correlation**: A functional relationship $g_\theta$ is constructed to map the offline prediction data $x_O(t)$ to the corresponding real-time measurement data $x_M(t)$. The parameters $\theta$ in $g_\theta$ (e.g., coefficients of a polynomial) are determined by minimizing the mean squared error between the mapped predictions and measurements over a time interval:

$$\theta^* = \arg\min_\theta \| g_\theta(X_O) - X_M \|^2, \tag{1}$$

where $X_O$ and $X_M$ represent the collections of corresponding data points within the fitting window. The optimized correlation $g_{\theta^*}$ maps the offline predictions of the cryogenic tank's thermodynamic state $x_O(t)$ into a data-informed prediction $x_D(t)$, which serves as an approximation of the corresponding real-time measurements $x_M(t)$:

$$x_M(t) \approx x_D(t) = g_{\theta^*}\big(x_O(t)\big). \tag{2}$$

This data-driven correlation refines the offline predictions, enabling them to more accurately reflect the actual system dynamics observed in real time.

- **Observation and Correction**: This is an event driven action, triggered either at the start of the experiment or when the prediction error exceeds a predefined threshold. When activated, the Monitoring Agent temporarily halts forecasting and waits for the collection of the measurement data $x_M(t)$, beginning at current time $t$ over a predefined window of duration $\Delta t_D$. The resulting measurement dataset of observation is:

$$X_{M,OB} = [x_M(t)\ \ x_M(t+\delta t)\ \ x_M(t+2\delta t)\ \ \dots\ \ x_M(t+\Delta t_D)]^\top, \tag{3}$$

where $\delta t$ is the sensor's data sampling interval. This newly collected dataset, along with the corresponding offline predictions $X_O$ over the same window ($[t, t+\Delta t_D]$), is used to fit new data-driven correlations. During the observation period, real-time forecasts are temporarily paused until the new data-driven correlations are ready.

- **Auto-Calibration**: It is a proactive, periodic, and non-disruptive process that updates the data-driven correlation using existing historical data already collected prior to the current time $t$. At the end of predefined intervals $\Delta t_{AC}$, the framework constructs the measurement dataset of the Auto-Calibration $X_{M,AC}$ over the data collection window of duration $\Delta t_D$ ending at the current time $t$:

$$X_{M,AC} = [x_M(t-\Delta t_D)\ \ \dots\ \ x_M(t-2\delta t)\ \ x_M(t-\delta t)\ \ x_M(t)]^\top. \tag{4}$$

This historical dataset $X_{M,AC}$ and its corresponding offline predictions $X_O$ ($[t - \Delta t_D, t]$) are used to update the correlation without collecting new data or interrupting ongoing real-time system state forecasting.

Figure 2 presents the conceptual workflow of the ARCTIC framework. Prior to mission execution or experimental operation, the cryogenic tank's geometric configuration, initial conditions, and boundary conditions expected during the mission are defined based on the system design parameters and mission specifications. These inputs are used to construct a nodal model, which generates offline predictions $x_O(t)$ of the tank's thermodynamic evolution over the planned mission. The offline predictions serve as a baseline and are later refined through data-driven correlations to enable accurate, real-time cryogenic tank thermodynamic state prediction under operational uncertainties.

As shown in Figure 2(a), at time $t = 0$, as the experiment begins, measurement data $x_M(t)$ begins streaming to the Monitoring Agent, and an initial observation period ($[0, \Delta t_D]$) is initiated. After this period, the Monitoring Agent fits the first set of data-driven correlations and starts to continuously generate data-informed forecasts $x_D(t)$ of key thermodynamic variables by demand (e.g., tank pressure and temperature) over a predefined time horizon $\Delta t_{AC}$ into the future.

The Monitoring Agent continuously compares data-informed forecasts with real-time sensor measurements. When the prediction error remains below a predefined threshold $\varepsilon$, determined based on the system's safety tolerance, the cryogenic tank system is assumed to operate under unchanged underlying physics. In this case, the historical sensor data collected so far are considered representative of the current thermodynamic behavior of the system and are therefore suitable for refining the offline predictions.

As shown in Figure 2(b), as new sensor data continues to stream in, the data-driven correlations are periodically updated to reflect minor fluctuations in the system state, incorporating the most recent system state. Specifically, if the error remains within the threshold $\varepsilon$ for a duration of $\Delta t_{AC}$, an Auto-Calibration action is triggered at the end of the duration. This action updates data-driven correlations using sensor data from the past interval $[t - \Delta t_D, t]$, where $t$ is the current time and $\Delta t_D$ is the length of data collection window. Auto-Calibration is performed periodically as long as the error remains below $\varepsilon$, ensuring that the correlations stay synchronized with the evolving system state and enabling the framework to continuously provide accurate forecasts over a prediction horizon of $\Delta t_{AC}$ ahead of the current time.

In contrast, if the Monitoring Agent detects that the discrepancy between predictions and real-time sensor data exceeds the threshold $\varepsilon$, it is interpreted as a significant change in the system's physical behavior, such as new event happening or a shift in boundary conditions. Under these circumstances, as shown in Figure 2(c), the framework will stop generating new forecasts. The existing data-driven correlations and all already collected sensor data are considered obsolete and no longer reliable for forecasting since they represent the system's state before the change. The Monitoring Agent then initiates an Observation and Correction action, as shown in Figure 2(d), during which new sensor data are collected over the interval $[t, t + \Delta t_D]$. Once the observation period concludes, new data-driven correlations are fitted using this updated dataset. These new correlations are used to correct the offline predictions and restore accurate, data-informed forecasting $\Delta t_{AC}$ into the future.

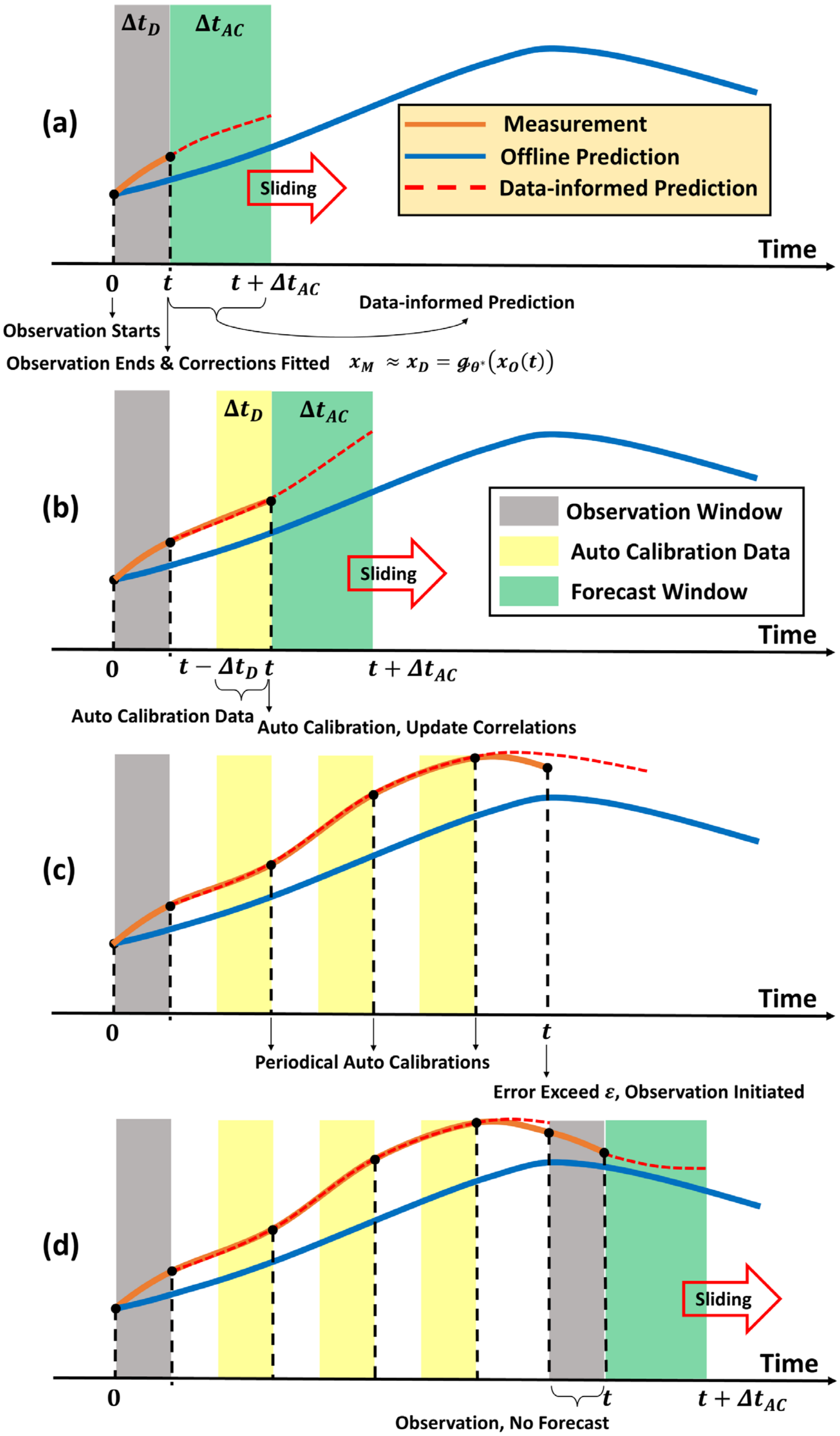

Figure 2. A conceptual workflow of ARCTIC illustrates forecasting, auto-calibration, and observation-and-correction. (a) As the experiment begins ($t = 0$), the Monitoring Agent collects measurements over the first observation window and fits the initial data-driven correlations to map offline predictions to sensor data. (b) When forecast errors remain below the threshold $\varepsilon$, the underlying physics is assumed unchanged. At the end of every auto-calibration interval $\Delta t_{AC}$, the Monitoring Agent updates the correlation using the most recent historical measurements in a sliding window of length $\Delta t_D$, without interrupting forecasting.

(c) If the discrepancy between forecasts and measurements exceeds $\varepsilon$, the framework interprets this as a physical or boundary-condition change. Forecasting is temporarily halted, and the current correlations are discarded. (d) A new observation period begins over $[t, t + \Delta t_D]$. Offline predictions and newly collected measurements within this window are used to refit the correlation. Forecasting resumes afterward with updated data-informed predictions.

Although sensor data are streamed continuously, updating the correlations too frequently may lead to instability due to random measurement noise. Thus, updates (Auto-Calibration) are performed at discrete intervals $\Delta t_{AC}$, allowing time for sufficient data accumulation and noise averaging. Auto-Calibration uses the most recent sensor data collected over the sliding window $[t - \Delta t_D, t]$, together with the corresponding offline predictions over the same window, ensuring that the updated correlation remains consistent with the baseline simulation while reflecting the current system state, especially when fluctuations exist in the boundary conditions. During the observation period $[t, t + \Delta t_D]$, the data-informed forecasts are temporarily unavailable. However, the offline simulation could continue to serve as a baseline estimation of the system state. The selection of $\Delta t_D$ and $\Delta t_{AC}$ is system-specific and case-specific, depending on how rapidly the system dynamics evolve. Fortunately, cryogenic storage tanks are typically slowly varying systems, making longer forecast horizons and infrequent updates sufficient. Several scenarios with varied settings of $\Delta t_D$ and $\Delta t_{AC}$ are demonstrated later in this study. Thus, through the coordinated use of Auto-Calibration and Observation and Correction, the Monitoring Agent ensures that predictions of the cryogenic tank's thermodynamic state remain accurate and reliable.

Owing to the non-intrusive architecture of the ARCTIC framework, the data-driven algorithm used for correlation fitting can be selected with considerable flexibility. Based on our previous work [26], the first-order regression is adopted in this study, as it provides an optimal balance between robustness and computational efficiency. This method is sufficiently fast for real-time applications and, when applied over the temporal window $\Delta t_D$, effectively filters out random sensor noise through inherent averaging. Consequently, the first-order regression is implemented as the data-driven algorithm within the Monitoring Agent in this work.

By directly mapping simulation outputs to sensor measurements, the data-driven corrections compensate not only for individual physical phenomena, but also for the combined modeling inaccuracies in the nodal simulation, including those in initial and boundary conditions, geometric simplifications, and unmodeled cryogenic fluid behaviors. As a result, the direct mapping approach offers a generalizable and efficient alternative to traditional parameter tuning, and the proposed framework provides a robust foundation for real-time thermodynamic state estimation in autonomous cryogenic management systems.

## 3. Model and Numerical Implementations of Cryogenic Tanks

This section presents the numerical models and empirical correlations used for nodal code simulations of cryogenic tanks.

### 3.1 Mass, momentum, and energy conservations

In system modeling, a fluid system is first divided into several distinct zero-dimensional control volumes, referred to as "lumps". Each lump contains mass and energy and is assumed to maintain a single equilibrium thermodynamic state, assuming perfect mixing of the fluid within the control volume [18]. Lumps follow the conservation laws of mass, energy and momentum. For Lump $i$ with mass $m_i$, the mass conservation equation is:

$$\frac{dm_i}{dt} = \sum_j \dot{m}_{ij} n_{ij}, \tag{5}$$

where $\dot{m}_{ij}(t)$ is the absolute value of the mass flow rate between Lump $i$ and its $j^{th}$ neighboring lump. The flow direction is denoted by $n_{ij}$, which equals either 1 or -1 based on the predefined positive flow direction. The energy conservation equation for the internal energy $U_i$ of Lump $i$ is:

$$\frac{dU_i}{dt} = \sum_j \dot{m}_{ij} * H_{i/j} n_{ij} + \sum_j k_{eff,ij}(T_j - T_i). \tag{6}$$

The first term on the right-hand side accounts for energy transfer via mass exchange between its adjacent lumps, where $H_{i/j}(t)$ is the specific donor enthalpy in either lump $i$ or the $j^{th}$ neighboring lump. The second term represents heat transfer between adjacent lumps and Lump $i$, where, $T(t)$ denotes the lump temperature, and $k_{eff,ij}$ is the effective conductivity between the target lump $i$ and its $j^{th}$ neighboring lump. The effective conductivity, $k_{eff,ij}$, includes both conduction and convection heat transfer and is typically determined using empirical correlations.

Lumps are interconnected through "inertial tubes", which are one-dimensional flow pipes to account for velocity and pressure drops. The momentum conservation equation for the Tube $i$ is given by:

$$\frac{d\dot{m}_i}{dt} = \frac{A_i}{L_i}\left(p_{up} - p_{down} + C_{1,i} + F_i \dot{m}_i |\dot{m}_i|^{C_{2,i}} + C_{3,i}\dot{m}_i^2 - \frac{K_i \dot{m}_i |\dot{m}_i|}{2\rho_{up} A_i^2}\right), \tag{7}$$

where $A_i$ and $L_i$ denote the cross-sectional flow area and the length of the tube, respectively. The pressures at the predefined upstream and downstream lumps are represented by $p_{up}(t)$ and $p_{down}(t)$. The coefficient $C_{1,i}$ represents the head coefficient due to body force, such as gravity or pump. $C_{3,i}$ is the recoverable loss coefficient caused by area and/or density change inside the tube. $K_i$ is also referred as the K-factor, which comes from the entrances, exits or shape change (bends, tees, etc.) induced pressure loss. In cryogenic tank simulations, $C_{1,i}$, $C_{3,i}$ and $K_i$ are usually set to 0. $F_i$ is the friction coefficient, and $C_{2,i}$ is dependent on the flow regime, which is 0 for laminar flow and nearly 1 for turbulence flow.

### 3.2 Heat and mass transfer across the interface

For two-phase fluid problems, a technique known as the twinned lump is utilized to simulate the fluid behavior near the liquid-vapor interface [18]. In this approach, a single lump is divided into two control volumes separated by an "imaginary interface", allowing each sub-volume to maintain its own thermodynamic state. The "imaginary interface" represents the physical liquid-vapor surface, with an equivalent heat and mass transfer area. It is assumed to be infinitely thin and maintained at the saturation temperature corresponding to the current pressure, enabling accurate modeling of interfacial phase-change processes.

The mass transfer due to phase change is calculated based on the energy jump conditions at the interface:

$$\dot{m}_{evap} = \frac{h_v A_{if}(T_v - T_{sat}) - h_l A_{if}(T_{sat} - T_l)}{h_{fg}}, \tag{8}$$

where $\dot{m}_{evap}(t)$ is the evaporation mass flow rate and $A_{if}$ is the interface area. The temperatures of the vapor, liquid, and saturation interface are denoted as $T_v(t)$, $T_l(t)$, and $T_{sat}(t)$, respectively. The heat transfer coefficients of the vapor and liquid are represented by $h_v(t)$ and $h_l(t)$, while $h_{fg}(t)$ is the latent heat of the phase change.

The simplest model for the interfacial heat transfer coefficients is the solid-like conduction model [18], which treats both the vapor and liquid lumps near the interface as fluid bulks with "solid-like" properties. In this model, the interfacial heat transfer coefficients ($h_l$ and $h_v$) are approximated by their respective thermal conductivities, $k$, to calculate heat transfer between the fluid bulks.

Since the interface in cryogenic tanks can be approximated as a flat plate, empirical correlations for natural convection over a flat plate can be employed to estimate the heat transfer coefficients under natural convection conditions [33]. In the liquid region, the heat transfer coefficient is:

$$h_l = 0.524 \frac{k}{L_l} \mathrm{Ra}^{1/5}, \tag{9}$$

where $L_l$ is the characteristic length of the liquid defined as the ratio of the interface area to the interface perimeter. The Rayleigh number can be calculated using the Grashof number Gr and the Prandtl number Pr:

$$\mathrm{Gr} = \frac{g\beta L_l^3 (T_{sat} - T_\infty)}{\nu^2},$$
$$\mathrm{Pr} = \frac{\mu c_p}{k}, \tag{10}$$
$$\mathrm{Ra} = \mathrm{Gr}\mathrm{Pr},$$

where $g$ is the gravitational acceleration, $\beta$ is the thermal expansion coefficient, and $\nu$ is the kinematic viscosity. The interface temperature and the bulk temperature are denoted by $T_{sat}$, and $T_\infty$, respectively. The dynamics viscosity is $\mu$, and $c_p$ is the specific heat.

For the vapor region, the Nusselt number is first determined using the following correlations [34]:

$$\begin{aligned} \mathrm{Nu}_v &= 4.5, & \mathrm{Ra}_v &\leq 10^7, \\ \mathrm{Nu}_v &= 0.08(\mathrm{Ra}_v)^{1/4}, & 10^7 < \mathrm{Ra}_v &\leq 10^{12}. \end{aligned} \tag{11}$$

After obtaining the Nusselt number, the interfacial heat transfer coefficient for the vapor region is calculated as

$$h_v = \frac{k \cdot \mathrm{Nu}_v}{L_v}. \tag{12}$$

### 3.3 Boundary layers of natural convection

Due to the zero-dimensional simplification and the perfect mixing assumption of the lumps, natural convection mixing effect and the boundary layers inside the cryogenic tank can only be modeled using empirical correlations. Treating the side walls of the cryogenic tank as vertical flat plates, the thickness of the boundary layer, $\delta$, is given by [35]:

$$\delta/x = 3.92\left(\frac{0.952 + \mathrm{Pr}}{\mathrm{Gr}_x \mathrm{Pr}^2}\right)^{1/4}, \qquad \text{Laminar}$$

$$\delta/x = 0.565\left(\frac{1 + 0.494Pr^{2/3}}{\mathrm{Gr}_x \mathrm{Pr}^{8/15}}\right)^{1/10}, \qquad \text{Turbulent} \tag{13}$$

where, $x$ represents the local elevation, which is the vertical distance from the bottom of the tank to the center of the lump in the liquid region. In the vapor region, $x$ is defined as the elevation from the interface to the center of the vapor lump. Consider the wall temperature $T_{wall,x}$ at elevation $x$, the non-dimensional numbers $Gr_x$ denotes the local Grashof number:

$$\mathrm{Gr}_x = \frac{g\beta x^3 \left(T_{wall,x} - T_\infty\right)}{\nu^2}. \tag{14}$$

The flow rates within boundary layers are given by [22]:

$$\dot{m} = 0.1666\pi R\rho\bar{v}\delta, \qquad \text{Laminar}$$

$$\dot{m} = 0.2872\pi R\rho\bar{v}\delta, \qquad \text{Turbulent} \tag{15}$$

where $R$ is the radius of the tank, $\rho$ is the density of the fluid. The average velocity $\bar{v}$ of the boundary layer is given by:

$$\bar{v} = 1.185\frac{\nu}{x}\left(\frac{\mathrm{Gr}_x}{1 + 0.494Pr^{\frac{2}{3}}}\right)^{1/2}. \tag{16}$$

**3.4 Sloshing**

Sloshing refers to the movement of liquid induced by external forces, such as mechanical vibrations, vehicle takeoff, and flight maneuvers. During sloshing, liquid motion circulates subcooled liquid near the liquid-vapor interface, leading to a significant increase in condensation and a corresponding pressure collapse. In the meanwhile, contact between the liquid and the hot wall due to sloshing can also cause rapid vaporization, resulting in a sudden pressure rise. These thermodynamic effects of sloshing are often critical to the safe operation of various mechanical systems [36]. In future space missions, thermodynamic effects due to sloshing may become even more significant due to extended mission durations and more abrupt vehicle maneuvers [32].

Due to the inherently three-dimensional nature of sloshing, developing accurate numerical models for nodal codes remain an area of active research. Current nodal simulations rely solely on empirical correlations derived from limited experimental data. Ludwig et al. [37] developed a set of empirical correlations for sloshing based on experiments conducted using a ground-based cryogenic tank. In their study, they developed an empirical correlation for the sloshing Nusselt number $\mathrm{Nu}_s$, to characterize the interfacial heat transfer effects of sloshing:

$$\mathrm{Nu}_s = \frac{\mathrm{Re}_s^{0.69}\mathrm{Pr}^{1/3}}{(\mathrm{Re}_s)_c^{0.69}} \tag{17}$$

The $\mathrm{Re}_s$ is the sloshing Reynolds number:

$$\mathrm{Re}_s = \sqrt{1.8}\,\eta \left(\frac{b}{R}\right)^2 \frac{(gR^3)^{1/2}}{\nu}, \tag{18}$$

where $\eta$ represents the frequency ratio, defined as the ratio of the sloshing frequency to the tank's natural frequency. The maximum amplitude of sloshing is denoted by $b$, while $R$ represents the tank's radius. The critical Reynolds number is given by $(\mathrm{Re}_s)_c$, which is the sloshing Reynolds number when $\mathrm{Nu}_s = 1$. Marques et al. [27] further calibrated the coefficients to better match the experiments. Therefore, the sloshing enhanced heat transfer coefficients are given by:

$$\begin{gathered} h_{v,s} = \frac{k_v}{R}\left(140\ \mathrm{Re}_{v,s}^{0.69}\mathrm{Pr}_v^{1/3}\right) \\ h_{l,s} = \frac{k_l}{R}\left(50\ \mathrm{Re}_{l,s}^{0.69}\mathrm{Pr}_l^{1/3}\right) \end{gathered} \tag{19}$$

## 4. Synthetic Test in Virtual Environment

To evaluate the capability of the ARCTIC framework in capturing the combined effects of multiple unmodeled phenomena, synthetic tests are conducted using virtual environment first. The cryogenic tank model, shown in Figure 3, is cylindrical with a height of 1 meter and an inner diameter of 1 meter. Initially, the tank contains a two-phase nitrogen mixture with a 75% liquid fill level, saturated at 100 kPa (77.24 K). The bottom and side walls are adiabatic, while heat flux is applied only to the tank's lid.

A Virtual Environment Model is used as a surrogate for experiments, simulating the thermodynamic state of cryogenic tanks. This model divides the tank vertically into ten lumps, with additional lumps representing boundary layers to account for natural convection effects. Empirical correlations for convection at a flat plate interface are incorporated to model the interfacial phase change. Meanwhile, a Simple Model is used to simulate the same tank but without using any empirical correlation and boundary layer lump. In this model, the interface heat transfer is treated as pure conduction, and boundary layers are not included. Table 1 summarizes the key differences between the Virtual Environment and Simple Models.

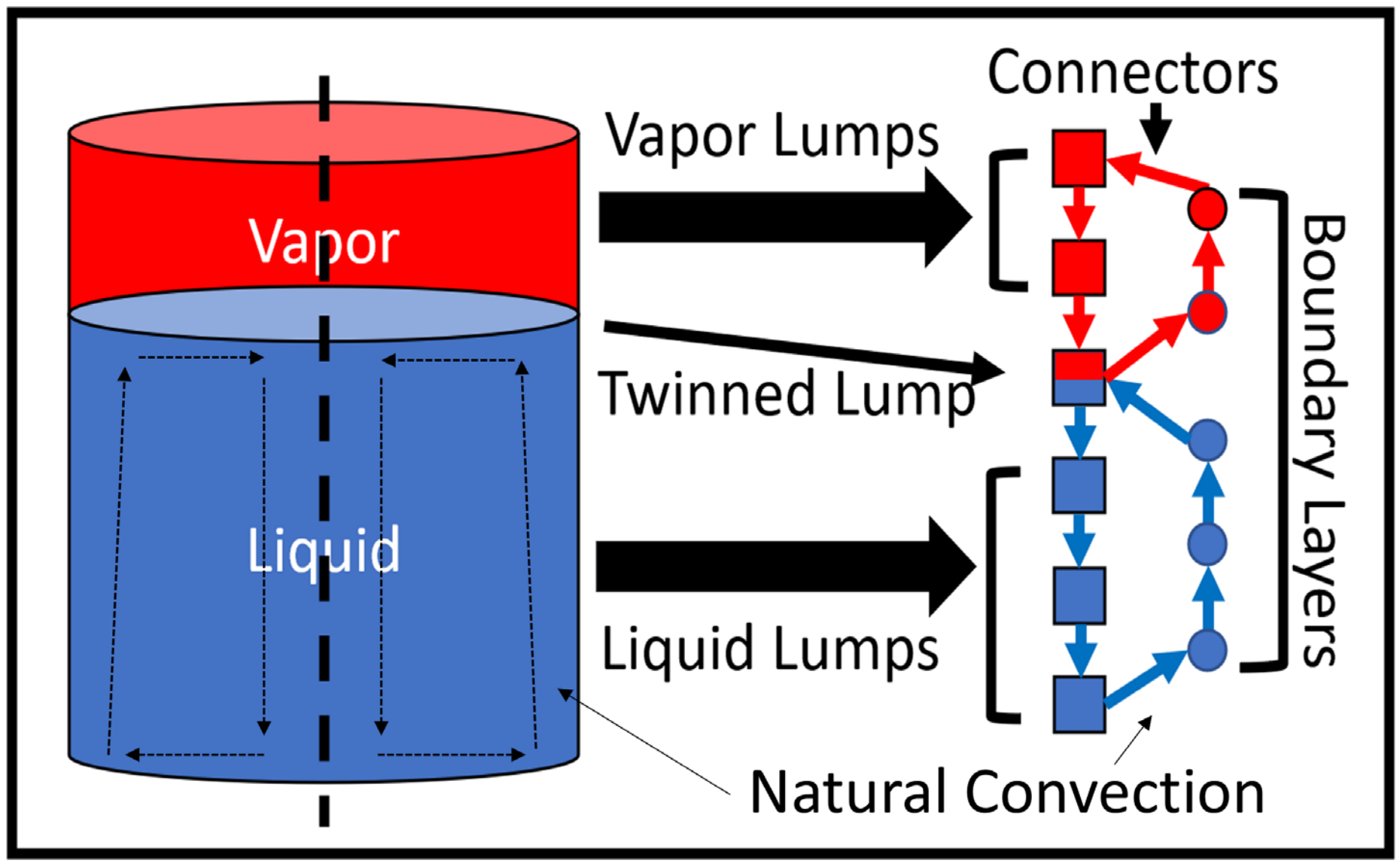

Figure 3. The Virtual Environment Model for a two-phase cryogenic tank. Due to the symmetry, only half of a two-phase cryogenic tank is modeled.

Table 1. Comparison of the Virtual Environment Model and the Simple Model.

| | **Virtual Environment Model (benchmark)** | **Simple Model** |
| --- | --- | --- |
| **Interface** | Natural convection [33,34] | Conduction |
| **Boundary layers** | Natural convection [22,35] | No |
| **Sloshing $h_s$** | Empirical correlation from Ludwig's [27,37] | Constant |
| **Data noise** | Gaussian noise | No |

In this section, simulation results from the Virtual Environment Model are treated as "measurement data" (denoted as Measurement), while outputs from the Simple Model are used as offline predictions. The Monitoring Agent then corrects the Simple Model predictions based on the Virtual Environment Model results, thereby compensating for the missing boundary layer effects and interfacial convection. To evaluate the robustness of the algorithm, boundary condition fluctuations are also introduced. Although long-term space missions typically span days to months, this section emphasizes the framework's performance under various events and operational scenarios. Therefore, the simulated flow time is limited to several minutes. The subsequent validation section will address the method's long-term performance.

To investigate the influence of data noise, Gaussian noise is added to the measurement data. The noise has a zero mean, and its standard deviation, $\sigma$, is set to 5% of the standard deviation of the original measurement data:

$$\sigma = 5\% \times std\big(x(t)\big) \tag{20}$$

Thus, the noise measurement data is:

$$x_{m,noise}(t) = x_m(t) + \tilde{x}(t), \qquad \tilde{x} \sim \mathcal{N}(0, \sigma) \tag{21}$$

**4.1 Long-term Self-pressurization with Fluctuations**

Cryogenic propellent tanks are typically well-insulated. However, there is still heat leakage from background radiation and structural contacts, leading to thermal stratification and self-pressurization. This effect is particularly critical for long-term missions, where accurately predicting the self-pressurization behavior of cryogenic storage tanks is essential for safety and CFM.

In this problem, a constant heat flux of $30W/m^2$ is applied to the tank's lid. As shown in Figure 4(d), a significant fluctuation in heat flux is introduced between $t = 150s$ and $t = 180s$ to mimic uncertainties encountered in real missions and to test the robustness of the real-time, data-informed simulations. The observation period is set to $\Delta t_D = 20s$, with a forecasting window of $\Delta t_{AC} = 40s$. The threshold of initiating observations for pressure prediction is set to $\varepsilon_p = 500$ Pa, corresponding to approximately 0.5% of the starting pressure. For interface temperature prediction, the threshold is set to $\varepsilon_{T,v} = 0.1$K.

Figures 4(a) and 4(b) present the ARCTIC forecasting of tank pressure and interface temperature, respectively. The interface temperatures are obtained from the twinned lump, in which a single lump is divided into liquid-side and vapor-side control volumes, with the imaginary surface between them representing the liquid-vapor interface. The temperatures of these two sub-volumes are used to represent the liquid-side and vapor-side interface temperatures, respectively. The results show that, after a short initial observation period, the real-time forecasts yield highly accurate predictions solely through auto-calibration.

Since no additional observations are triggered, and auto-calibration introduces no time delays, ARCTIC consistently provide accurate forecasts of $\Delta t_{AC}$ ahead of real-time, enabling proactive decision-making immediately after the first observation. During the observation window, indicated by the gray shaded region, forecasting is stopped. The offline prediction within this region represents the fitted results obtained after the correlation has been obtained. Figure 4(c) illustrates the impact of data noise, showing that the Monitoring Agent effectively derives data-driven correlations despite the presence of noise. Considering the effects of data noise, the threshold is relaxed to 2% of the initial pressure. These findings demonstrate the robustness of the ARCTIC framework for real-time cryogenic tank forecasting under realistic and noisy conditions.

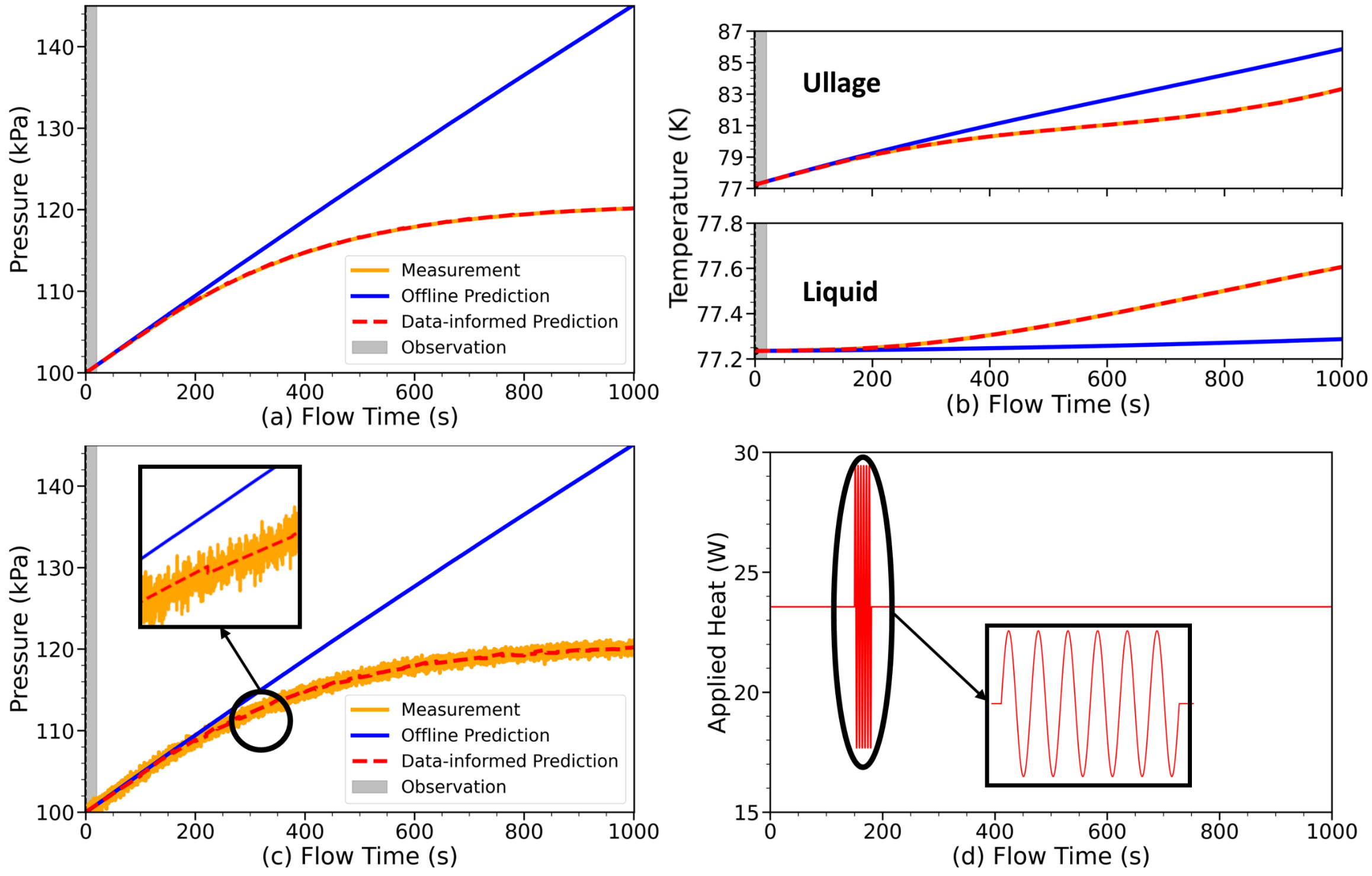

Figure 4. (a) Comparison of offline predictions, ARCTIC forecasts, and clean measurement data for tank pressure. (b) Comparison of offline predictions, ARCTIC forecasts, and clean measurement data for ullage (vapor) and liquid temperatures near the interface. (c) ARCTIC forecasts of tank pressure using noisy measurement data. (d) Applied heat flux boundary condition at the tank lid. A sinusoidal fluctuation is introduced between 150 s and 180 s to simulate transient disturbances.

### 4.2 Pressurization and Holding

The preparation of the cryogenic tank typically involves liquid filling, liquid level adjustment, gas injection (pressurization), and holding. During the holding stage, the external heat input becomes negligible compared to the violent pressurization caused by gas injection. With the continuous internal mixing and heat transfer between the hot gas and cold liquid, the tank pressure gradually decreases until reaching a stable thermodynamic state. Figure 5 (d) shows the boundary conditions used in this holding problem: a heat load of 39.27 $W$ is applied to the tank's lid for the first 500 seconds, after which the heat load is set to zero, and the tank is held for another 500 seconds. The observation period is set to $\Delta t_D = 20s$, with a forecasting window of $\Delta t_{AC} = 40s$. The thresholds for pressure and temperature are selected consistently with the values defined in Section 4.1.

Figures 5(a) and 5(b) present the ARCTIC forecasting for the tank pressure and the temperature near the interface, respectively. The results indicate that shortly after the tank transitions from the pressurization to the holding, the Monitoring Agent initiates a new observation. As previously discussed, this is triggered by a significant discrepancy between the measurement and the data-informed predictions, which the Monitoring Agent interprets as a change in the underlying physical behavior. In response, it discards prior data and fits new data-driven correlations. Following this second observation, the data-informed predictions maintain high accuracy using auto-calibration only until the end of the simulation. Although external heating ceases during the holding phase, heat transfer from the hot gas to the liquid near the interface continues. Consequently, the transition from pressurization to holding has a minor impact on liquid temperature evolution, as shown in Figure 5(b), and one single observation is sufficient for accurate data-informed temperature prediction. Figure 5(c) demonstrates the impact of measurement noise. Due to the relaxed threshold, the effects of the transient are fully captured by auto-calibration, and no additional observation is triggered. These results confirm that the ARCTIC algorithm is robust, capable of maintaining accurate data-informed predictions even in the presence of measurement noise.

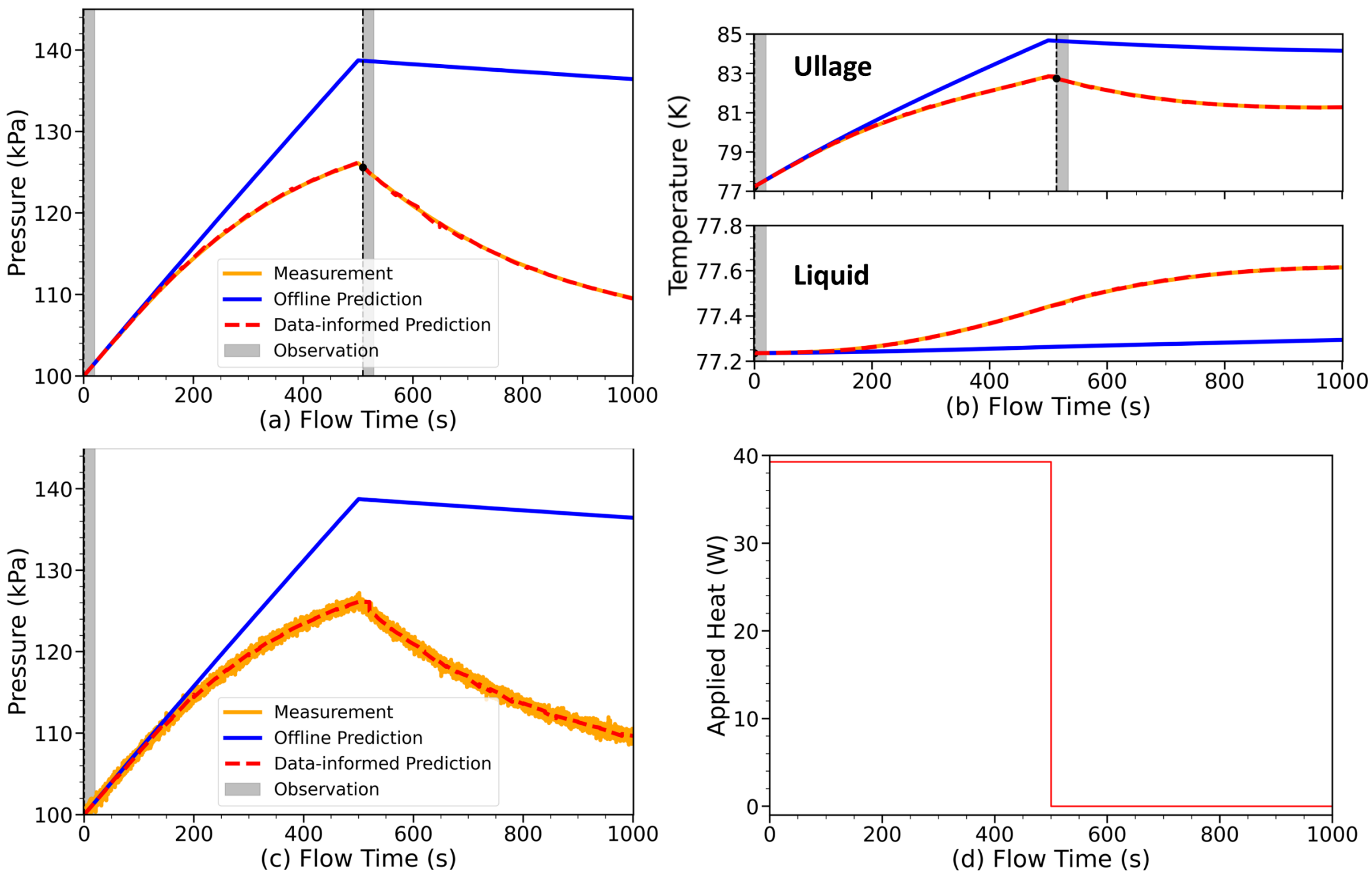

Figure 5. (a) Comparison of offline prediction, ARCTIC forecasts, and clean measurement data for tank pressure. (b) Comparison of offline prediction, ARCTIC forecasts, and clean measurement data for ullage (vapor) and liquid temperatures near the interface. (c) ARCTIC forecasts of tank pressure using noisy measurement data. (d) Applied heat flux boundary condition at the tank lid. The heat flux is set to zero during the holding phase, which begins at t = 500 s.

### 4.3 Sequential Events

A sequential multi-event problem is used to evaluate the performance of data-informed predictions in a dynamically changing environment. As shown in Figure 6(d), a small heat load is initially applied to the tank for 450 seconds, with a fluctuation occurring between $t = 150s$ and $t = 200s$. The heat load then increases significantly for another 350 seconds. After $t = 800s$, the external heat load becomes negligible, and the tank transitions into a holding stage. The observation period is set to $\Delta t_D = 20s$, with a forecasting window of $\Delta t_{AC} = 40s$ and the same threshold selections.

Figures 6(a) and 6(b) present the ARCTIC forecasting for tank pressure and temperature near the vapor/liquid interface, respectively. The results indicate that variations in external heat flux have a less pronounced effect than the transition from self-pressurization to holding on both tank pressure and vapor temperature. As a result, no new observation is initiated around $t = 450s$. For liquid temperature, since heating primarily occurs through heat transfer from the hot gas at the interface, the data-driven correlation remains almost unaffected by the changing boundary conditions, requiring only a single observation. Figure 6(c) illustrates the impact of data noise. The noise broadens the original measurement curve, increasing the discrepancy between data-informed predictions and measurements. However, with the relaxed threshold, one new observation is still sufficient as the clear data. This result suggests that for high-noise scenarios, increasing the tolerance threshold based on the noise magnitude improves prediction's efficiency.

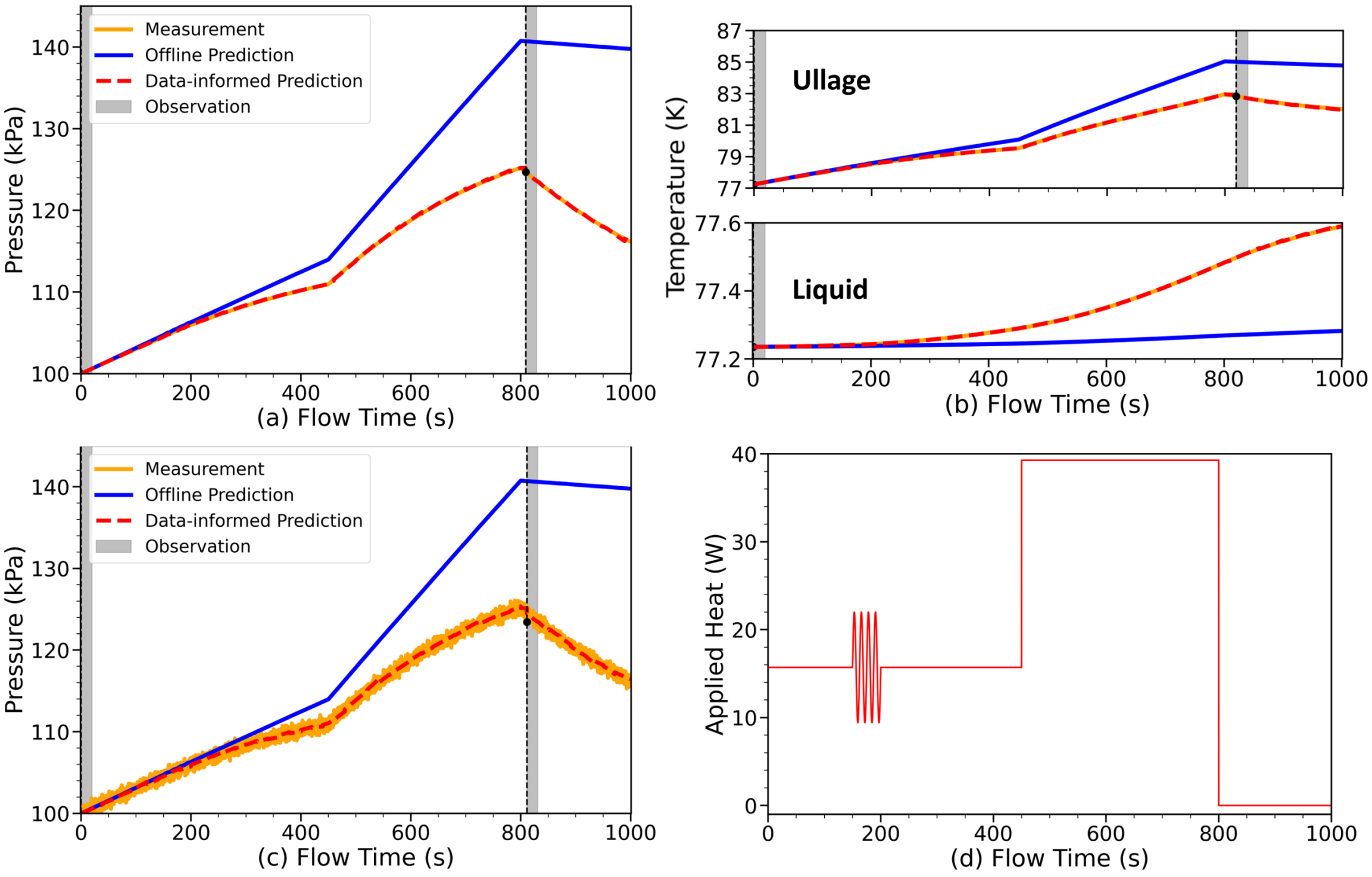

Figure 6. (a) Comparison of offline prediction, ARCTIC forecasts, and clean measurement data for tank pressure. (b) Comparison of offline prediction, ARCTIC forecasts, and clean measurement data for ullage (vapor) and liquid temperatures near the interface. (c) ARCTIC forecasts of tank pressure using noisy measurement data. (d) Applied heat flux boundary condition at the tank lid. A sinusoidal fluctuation is introduced between 150 s and 200 s, and step changes in the applied heat flux occur at t = 450 s and t = 800 s.

### 4.4 Periodical Operated (Mixing and Venting)

Periodic venting and mixing are among the most common methods for cryogenic tank pressure control. While recent research has increasingly focused on zero-boil-off (ZBO) strategies [38] to minimize cryogenic liquid losses, understanding the complexities of periodic operations remains essential for improving nodal simulation capabilities, as such operations present significant challenges to traditional nodal models. Moreover, accurate simulation of venting processes is critical for future on-orbit refueling applications.

In this section, a periodic venting and mixing scenario is examined. Following an initial 200 seconds of pressurization with a constant heat flux of $20W/m^2$, the periodic operation starts. As illustrated in Figure 7(d), venting occurs through a valve on the tank lid at a constant mass flow rate, lasting for 50 seconds every 100 seconds. To maintain mass conservation, cold liquid is injected from the bottom at the same mass flow rate, thereby inducing strong convection through enhanced mixing. Given the short venting duration and rapid transients involved, the observation period is set to $\Delta t_D = 20s$, with a forecasting window of $\Delta t_{AC} = 20s$. It is worth noting that the venting period adopted in this case is shorter than those typically observed in cryogenic tank experiments [26] and is intentionally chosen to represent an extreme case for testing the robustness of the proposed framework.

Figure 7(a) presents the pressure evolution of the cryogenic tank. During each cycle, the tank pressure decreases over a 50-second period when the venting valve is open, then rises again once the valve is closed. Consequently, the tank pressure evolution curves exhibit recurring cycles that are similar in shape but not identical. After the first observation, the data-informed simulation relies solely on auto-calibration to provide continuous, accurate 20-second forecasts. Figure 7(b) shows the vapor and liquid temperature evolutions. The data-informed predictions for both temperature curves demonstrate similar behavior to pressure predictions, maintaining high accuracy through auto-calibration. Because the pressure changes in this case are significantly smaller than in previous cases, the noise threshold is reduced to 1% of the initial pressure. The corresponding results are shown in Figure 7(c). The slightly unsmooth and discontinuous features observed in the prediction curves reflect the periodic updates of the data-driven correlations via auto-calibration.

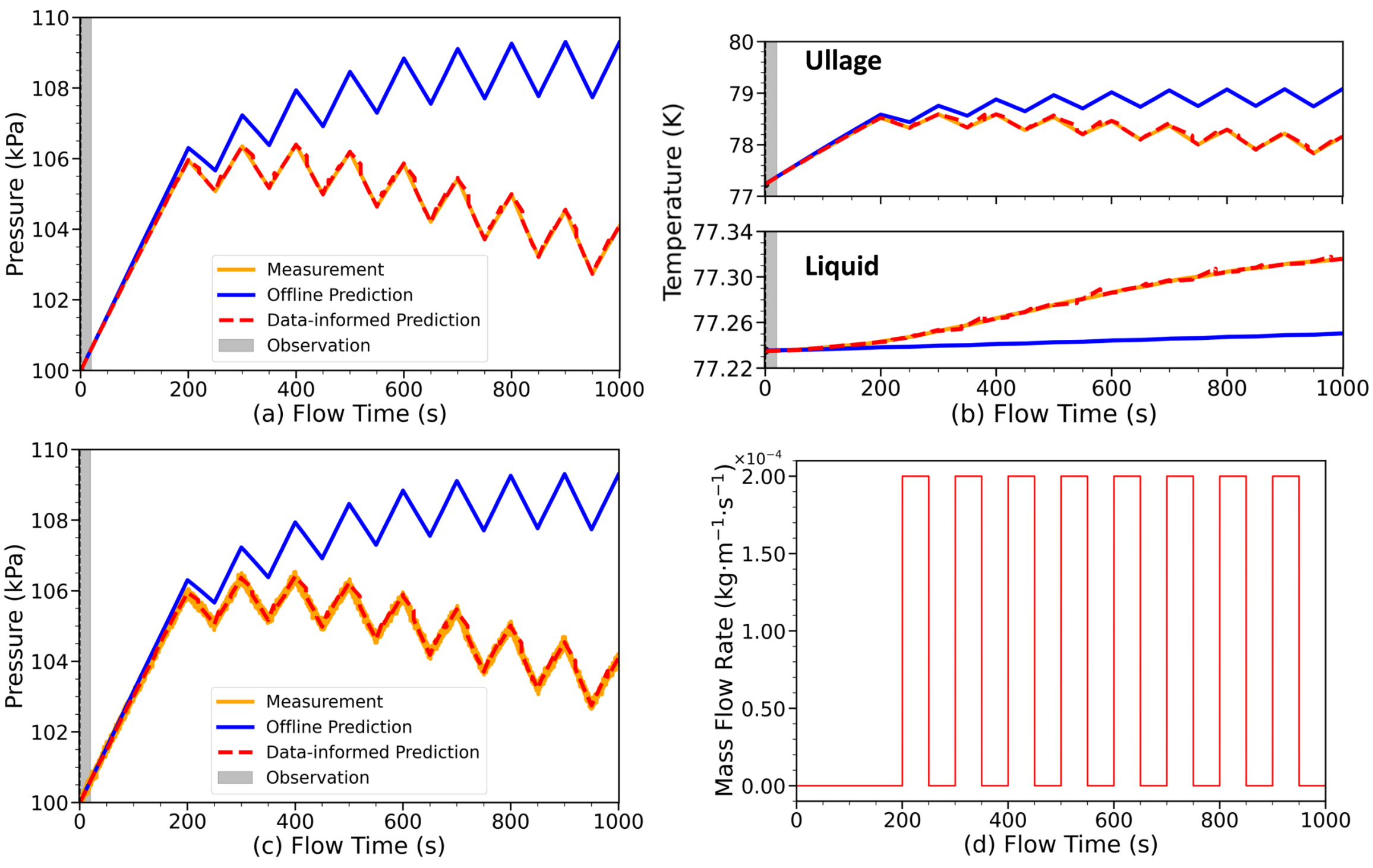

Figure 7. (a) Comparison of offline prediction, ARCTIC forecasts, and clean measurement data for tank pressure. (b) Comparison of offline prediction, ARCTIC forecasts, and clean measurement data for ullage

(vapor) and liquid temperatures near the interface. (c) ARCTIC forecasts of tank pressure using noisy measurement data. (d) Mass flow rates at the tank inlet and outlet. Periodic operation begins at t = 200 s.

### 4.5 Sloshing

Sloshing is a fast transient event induced by external forces, often leading to violent pressure changes that pose safety risks. Due to its inherently three-dimensional nature and the complex deformation of the liquid-vapor interface, nodal simulations face significant challenges in accurately capturing sloshing dynamics. Moreover, experimental data and empirical correlations for sloshing are also limited, particularly under microgravity conditions.

In this case, after 100 seconds of self-pressurization with a heat flux of $100W/m^2$, a moderate sloshing event is introduced at $t = 100s$, characterized by $\eta = 0.8$ and $b/R = 0.12$. It is worth noting that, $b/R$ is treated here as a single control parameter, and, based on the observations reported in [37], the chosen value is well below the onset of breaking, which occurs at $b/R = 0.54$. This event lasts for 30 seconds and causes a dramatic increase in the interfacial heat transfer coefficient by three orders of magnitude, as illustrated in Figure 8(d). Given the fast transient nature of this problem, the observation period is set to $\Delta t_D = 4s$, with a forecasting window of $\Delta t_{AC} = 4s$.

Figure 8(a) shows a sharp pressure drop immediately following the sloshing event. As expected, the Monitoring Agent promptly initiates a new observation in response. After the observation and correction, ARCTIC resumes real-time forecasting and accurately tracks the pressure evolution until the end of the problem. Figure 8(b) presents the temperature evolution in the vapor and liquid regions near the interface. After sloshing occurs, intense interfacial heat and mass transfer drive the vapor and liquid towards nearly the same temperature. A new observation is triggered to capture the effects of sloshing, and the updated data-driven correlations remain valid throughout the rest of the simulation. Figure 8(c) illustrates the impact of measurement noise, demonstrating that ARCTIC remains robust and maintains accuracy even in the presence of noise for fast transient events.

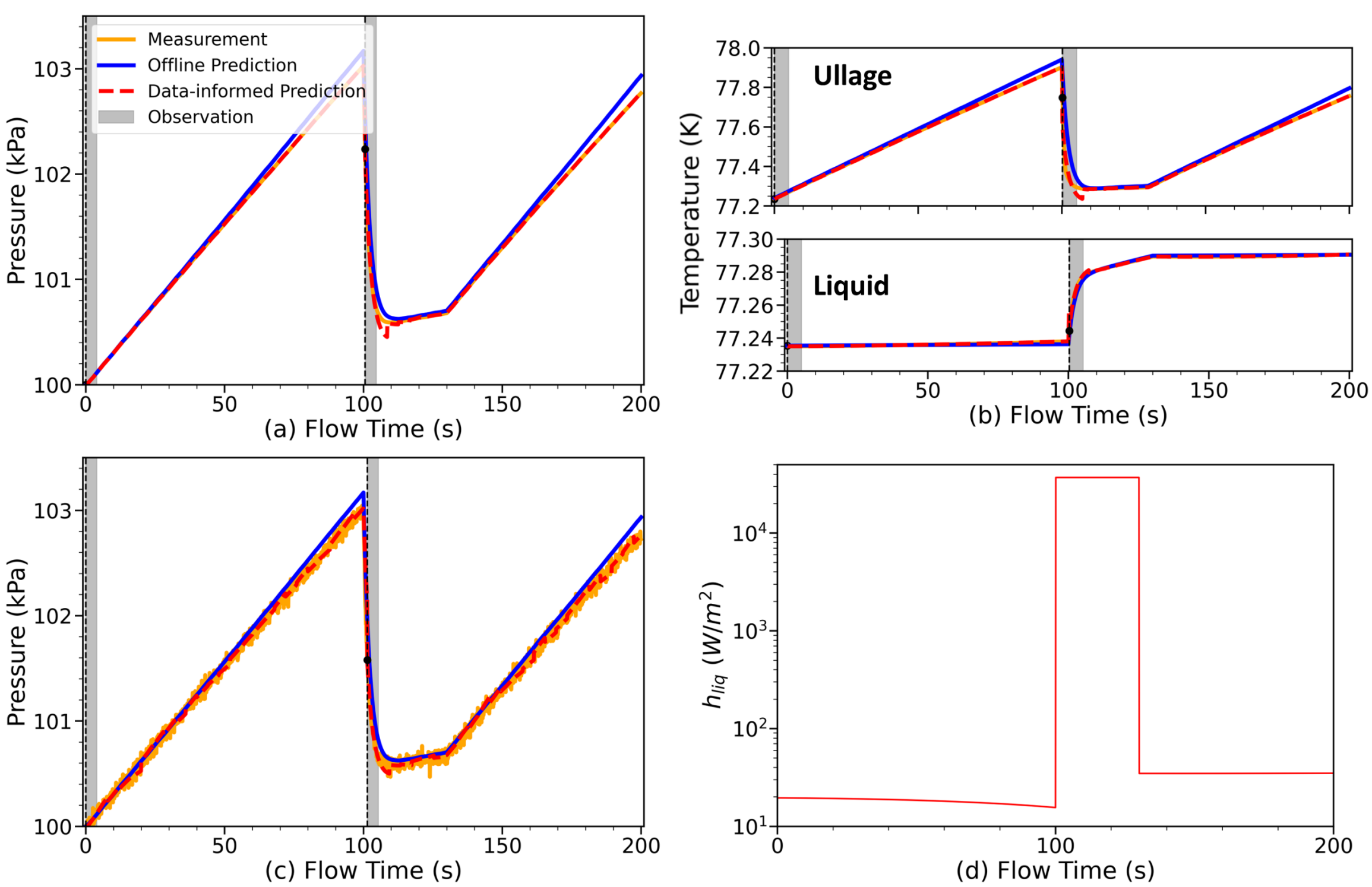

Figure 8. (a) Comparison of offline prediction, ARCTIC forecasts, and clean measurement data for tank pressure. (b) Comparison of offline prediction, ARCTIC forecasts, and clean measurement data for ullage (vapor) and liquid temperatures near the interface. (c) ARCTIC forecasts of tank pressure using noisy measurement data. (d) Interface heat transfer coefficients on the liquid side. A moderate sloshing event occurs at t = 100 s and lasts for 30 s, leading to a significant increase in the interface heat transfer coefficient.

## 5. Experiment Validations

In this section, cryogenic tank experiments conducted at several NASA research centers are used to validate the ARCTIC framework. The nodal models used in this section incorporate empirical correlations as the Virtual Environment Model to enhance the accuracy of offline predictions. The experimental conditions are summarized in Table 2. The details and validation results of each experiment are discussed individually in the following sections.

Table 2. Summary of Primary Experimental Conditions for Validation Cases

| | **Liquid** | **Experiment** | **Initial Pressure** | **Volume** | **Fill Level** | **Flow Time** |
|---|---|---|---|---|---|---|
| **MHTB [28]** | $LH_2$ | Self-Pressurization | 111.6 kPa | $18.09\ m^3$ | 50% | 850 min |
| **K-Site Tank 1 [30]** | $LH_2$ | Self-Pressurization | 103 kPa | $4.89\ m^3$ | 50% | 65000 s |
| **K-Site Tank 2 [32]** | $LH_2$ | Sloshing | 14.6 psi | 1750 L | 64% | 200 s |

### 5.1 Multipurpose Hydrogen Test Bed Self-Pressurization Experiments

The Multipurpose Hydrogen Test Bed (MHTB) is a large-scale experimental facility at NASA's Marshall Space Flight Center to study cryogenic propellant storage and management under low Earth orbit (LEO) conditions [28]. The MHTB tank, as shown in Figure 9(a), is cylindrical, with a diameter and height of $3.05\mathrm{m}$, featuring two 2:1 elliptical end caps. It has an internal volume of $18.09\mathrm{m}^3$ and a surface area of $35.74\mathrm{m}^2$, representing a full-scale liquid hydrogen ($LH_2$) tank. Multiple experiments have been conducted using the MHTB, including self-pressurization studies and spray bar thermodynamic vent system tests [28]. Numerical simulation models, including CFD models [29] and system-level models [21], have also been developed and validated against experimental data from MHTB tests.

In this section, experimental data from the 50% fill level self-pressurization experiment is used to validate the ARCTIC framework. The MHTB tank is filled with liquid hydrogen to 50% fill level. The initial pressure is 111.6 kPa, and the tank reaches saturation temperature at the start of the experiment. According to a previous study [39], the heat flux is $0.89873\mathrm{W/m}^2$ in the vapor region and $2.0841\mathrm{W/m}^2$ in the liquid region. The dynamic process is simulated over a total duration of 850 minutes. Due to limitations in available experimental data, only pressure evolution is analyzed. The observation period is set to $\Delta t_D = 10\ min$, with a forecasting window of $\Delta t_{AC} = 10\ min$.

Figure 9(b) presents the results of the data-informed simulations. After the first 10 minutes observations, the data-informed simulations provide 10 minutes of predictions continuously for the rest of the problem, significantly improving the accuracy of the nodal simulations.

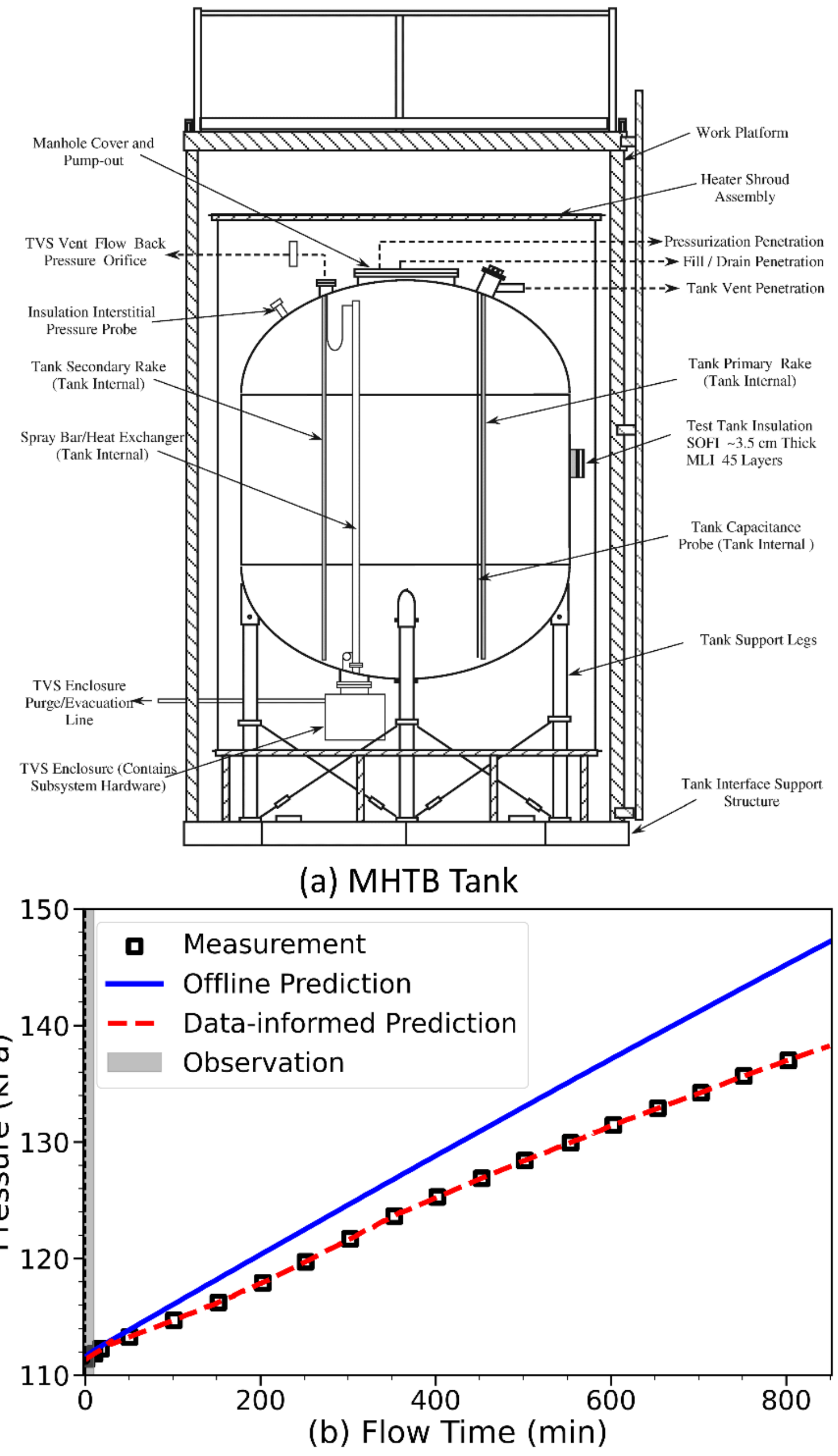

Figure 9. (a) Illustration of the MHTB tank [28]. (b) Comparison of the offline prediction, ARCTIC forecast, and the experiment data of tank pressure.

### 5.2 K-Site Tank Self-Pressurization Experiments

The K-Site tank pressurization experiment was conducted using a flight-weight, cryogenic liquid hydrogen storage tank at the K-Site facility in NASA's Glenn Research Center (formerly Lewis Research Center) [30]. As shown in Figure 10(a), the tank has an approximately ellipsoidal shape with a major-to-minor axis ratio of 1.2, a major diameter of 2.2m, and a total volume of $4.89\mathrm{m}^3$. A series of self-pressurization experiments were performed on the K-Site tank under various fill levels and applied heat fluxes [30]. In this analysis, we focus on the case with a 50% fill level and an applied heat flux of $3.5\ \mathrm{W/m^2}$ to validate the ARCTIC framework. The K-Site tank features a wall with varying thickness. However, due to the unavailability of detailed wall thickness data, the tank wall is assumed to have zero thickness in the model. This simplification will lead to an overestimation of the pressurization rate, as suggested by Kartuzova et al. [31]. The total simulated flow time is 65,000 seconds. The observation period is set to $\Delta t_D = 10\ min$, with a forecasting window of $\Delta t_{AC} = 10\ min$.

Figure 10(b) presents the data-informed simulation results for tank pressure, while Figures 10(b) and 10(c) show the temperature evolution curves at sensor 8 and sensor 16, respectively, as marked in Figure 10(a). The results from all three data-informed simulations demonstrate that the proposed approach can effectively rely on a short observation period and auto-calibration to accurately and continuously predict the K-Site tank's thermodynamic state with a 10-minute forecasting window. This predictive capability provides a valuable time window for potential control actions.

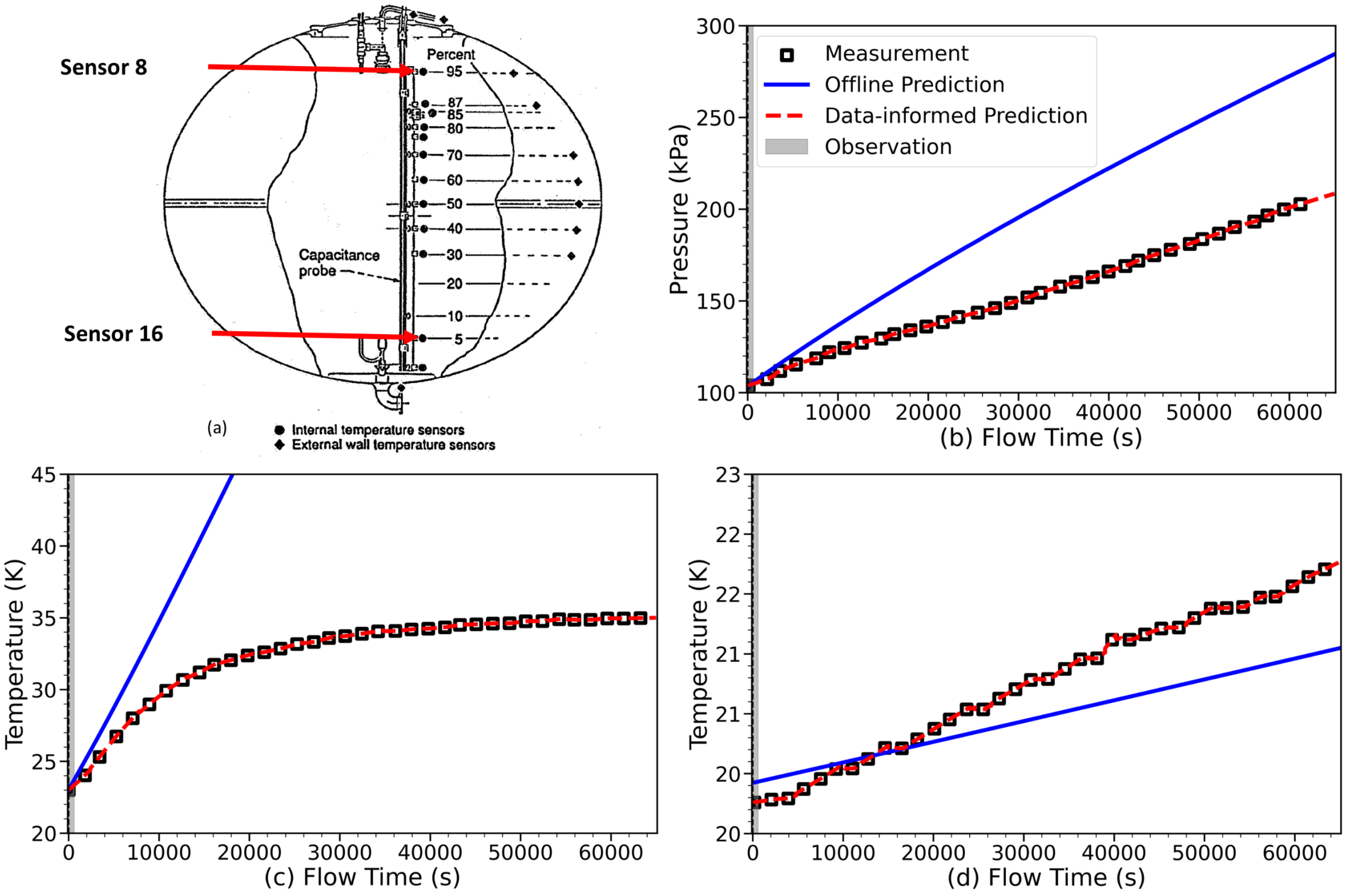

Figure 10. (a) Illustrations of K-Site tank for self-pressurization experiments and the locations of temperature sensors [30]. (b) Comparison of the offline prediction, ARCTIC forecasts, and the experiment data of tank pressure. (c) Comparison of the offline prediction, ARCTIC forecasts, and the experiment data of vapor temperature at sensor 8. (d) Comparison of the offline prediction, data-informed prediction and the experiment data of liquid temperature at sensor 16.

### 5.3 K-Site Tank Sloshing Experiments

A spherical $LH_2$ tank was added to the K-Site facility in 1990 [32] to study the thermodynamic response of the system during sloshing under normal gravity conditions. As shown in Figure 11(a), this tank has a spherical shape with a volume of 62 cubic feet ($1.756\ m^3$). Slosh frequency, amplitude, and ullage volume were systematically varied to investigate their effects on tank pressure and temperature [32]. In this section, sloshing experiment #870 is used to validate the ARCTIC framework. Experiment #870 is relatively well-documented and was previously analyzed by Kartuzova et al. using CFD simulations [40]. The initial pressure of the tank was 14.6 psi (100.67 kPa), with a 64% fill level. At the beginning of the experiment, as shown in Figure 11(b), hydrogen gas with the temperature of 41K was injected into the tank to pressurize it until the internal pressure reached 35.5 psi (244.76 kPa). The mass flow rate was then gradually reduced to maintain constant tank pressure. The valve was closed after holding and sloshing was initiated. The sloshing amplitude increased over time, reaching its maximum at $\mathrm{t} = 58$ s, after which it remained constant for the rest of the experiment.

Due to limitations in available experimental data, certain details, such as the exact mass flow rate changing curve, are unknown. Therefore, assumptions and simplifications were applied in the offline simulations:

1. As shown in Figure 11(b), the injection mass flow rate is assumed to increase linearly until reaching its maximum, then remains constant throughout the pressurization phase. The total injected hydrogen mass matches the experimental report.
2. During the holding period, instead of dynamically adjusting the mass flow rate to maintain constant pressure, the offline simulation assumes a constant mass flow rate, ensuring the total injection amount matches the experimental data.
3. For sloshing conditions, the offline simulation neglects the gradual amplitude increase, assuming zero sloshing until $\mathrm{t} = 58$ s, at which point the maximum amplitude is instantly reached and held constant.

The discrepancies between the experimental conditions and offline simulations highlight the necessity of ARCTIC framework to refine predictions and correct simulation errors caused by these uncertainties. The observation period is set to $\Delta t_D = 3\ s$, with a forecasting window of $\Delta t_{AC} = 2\ s$.

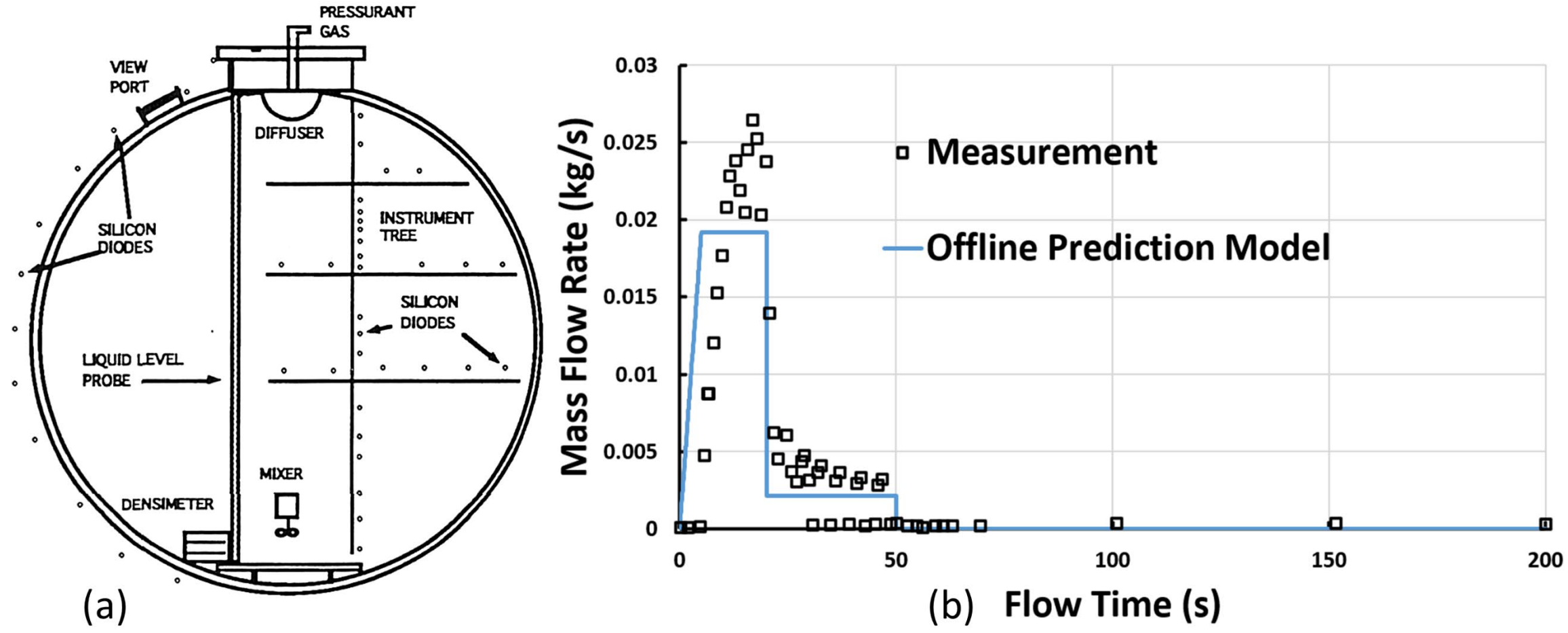

Figure 11. Illustrations of the K-Site tank for sloshing experiments [32]. (b) Mass flow rate of the injected gas during the experiment #870.

Figure 12 presents the results of the data-informed simulations. Due to discrepancies in the gas injection conditions, the offline predictions fail to accurately capture the constant tank pressure during the holding period and underestimate the pressure collapse caused by sloshing. In contrast, the data-informed predictions, based on only two observations, closely match the experimental data points. Although the data-informed predictions exhibit some fluctuations due to experimental noise, they significantly improve the offline predictions. The results demonstrate that, for most of the experiment's duration, the data-informed approach successfully predicts tank pressure for 2 seconds ahead of the current flow time, providing valuable information for the control system. A sensitivity study for different combinations of observation period $\Delta t_D$ and forecasting window of $\Delta t_{AC}$ is given in Appendix II.

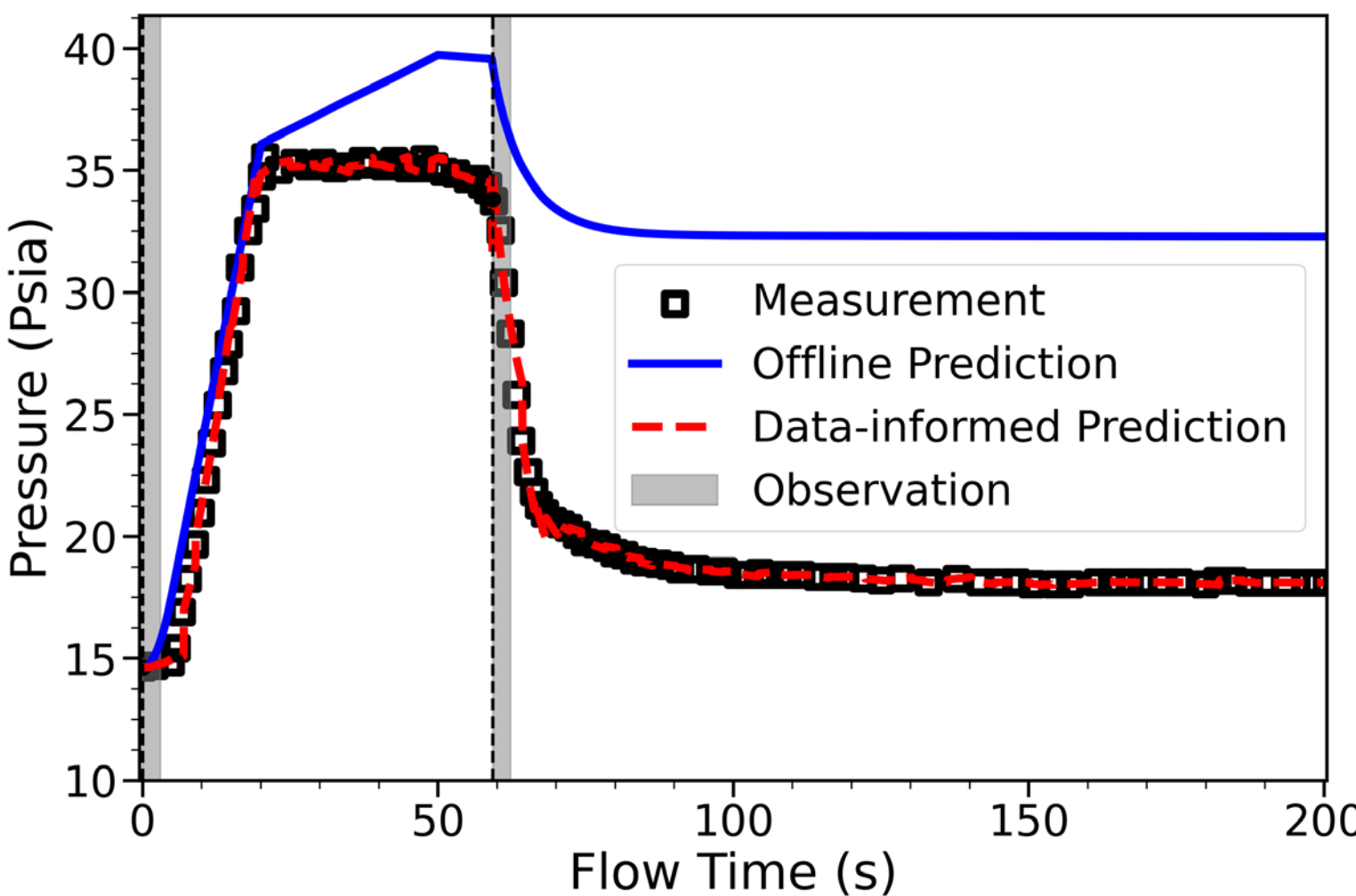

Figure 12. Comparison of the offline prediction, ARCTIC forecasts, and the measurement data of tank pressure.

**6. Demonstration of Adaptive Forecasting and Control Under Offline Model Deviation**

The previous sections demonstrated that the proposed ARCTIC framework can generate accurate forecasts based on offline simulations, which accounts for pre-scheduled control actions, such as increasing the venting mass flow rate. However, another practical scenario could involve cases where necessary control actions are not planned during offline simulations due to the inaccurate offline simulation, and these control actions are initiated on-the-fly during tank storage period based on the ARCTIC predictions. Therefore, this section investigates whether the proposed framework can still provide reliable forecasts when new control actions are initiated and are not included during the offline simulation stage.

To evaluate this capability, we present a synthetic test case as a demonstration of adaptive pressure control. In this scenario, the offline simulation underestimates the cryogenic tank's internal pressure due to an underestimated external heat leakage, representing, for instance, increased background radiation during the experiment. During the simulated experiment, the data-informed prediction indicates that the pressure will exceed a predefined safety threshold. In response, the autonomous control system schedules and executes a new venting action that was not part of the original offline simulation plan.

The tank configuration used in this case is identical to that of Section 3, with heat flux applied through the tank lid. The offline prediction is generated using the Virtual Environment Model with an applied heat flux of $10\ \mathrm{W/m^2}$, while the synthetic measurement data are generated with a higher heat flux of $15\ \mathrm{W/m^2}$. As shown in Figure 13(a), the offline prediction significantly underestimates the internal pressure due to the underestimated heat input. However, the data-informed prediction, with the observation period of $\Delta t_D = 5s$ and the forecasting window of $\Delta t_{AC} = 50s$, successfully detects the deviation and anticipates the over-pressurization event 50 seconds in advance. Therefore, at $t = 150s$, the control system schedules a venting action at $t = 200s$.

As shown in Figure 13(b), when the venting action begins at $t = 200s$, a discrepancy arises between the data-informed prediction and the real-time sensor measurements, triggering a new 5-second observation period to update the correlation. After the venting concludes at $t = 250s$, another observation is initiated to re-fit the data-driven correlations based on the new system behavior. The results demonstrate that, despite the offline simulation not accounting for the unscheduled control event, the proposed framework successfully corrected the predictions in real time. The data-informed predictions maintain an accurate 50-second forecast horizon and are effectively re-synchronized with the measurements after the new event.

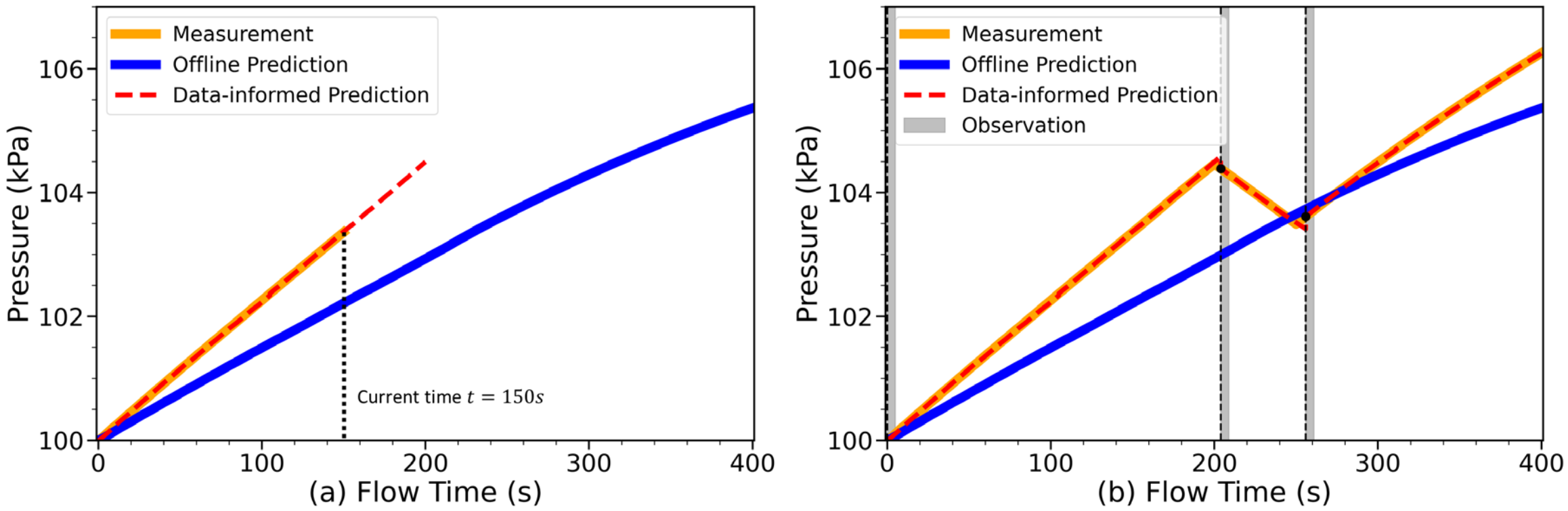

Figure 13. (a) Due to the underestimated external heat flux, the offline prediction underestimated the tank pressure. At $t = 150s$, the data-informed prediction gives a pressure prediction at $t = 200s$, indicating a new venting action is necessary. (b) Comparison of the offline prediction, ARCTIC forecasts, and the measurement data of tank pressure with the new venting action.

This demonstration illustrates how ARCTIC operates together with a cryogenic tank control system and highlights the physical significance of this capability. In a control context, ARCTIC enables predictive autonomy by using data-informed forecasts to anticipate the tank's thermodynamic response to changing boundary conditions. The control system can rely on the data-informed predictions produced by ARCTIC to schedule corrective actions, such as the venting action shown in Figure 13, before critical thresholds are reached. When a new event occurs that was not represented in the offline simulation, ARCTIC initiates an observation period to collect representative data and establish a new mapping that reflects the updated system behavior. Physically, the offline nodal simulation represents the expected nominal response defined prior to the mission, whereas real-time operation inevitably introduces deviations due to unexpected boundary conditions, unplanned control actions, or other unmodeled events. The data-driven mapping in ARCTIC functions as an adaptive correction layer that continuously aligns the offline model with the evolving thermodynamic state of the tank. By learning the most recent correlations between prediction and measurement, ARCTIC maintains the predictive validity of the digital model and ensures that the control system remains synchronized with the true system dynamics.

For more complex scenarios in which the system changes substantially after a newly scheduled control action, such as periodic venting and mixing, re-simulation of the system may become unavoidable. Under these conditions, ARCTIC can be extended with an additional routine that automatically launches a new simulation whenever a control action not included in the original offline predictions is scheduled based on ARCTIC's forecast of the future thermodynamic state. In this routine, the current data-informed prediction serves as the initial condition for the new simulation, and the updated boundary conditions incorporate the newly scheduled control action. ARCTIC then treats the output of this new simulation as the updated offline prediction and continues generating real-time data-informed forecasts. However, in such cases, the performance of the control system becomes limited by the computational cost of re-running numerical simulations, and system-level nodal codes may not be fast enough to support repeated re-simulations. For these scenarios, neural-network-based surrogate models may offer a promising alternative, providing the fast re-simulation capability needed for ARCTIC to operate effectively in onboard cryogenic tank control applications [41].

## 7. Conclusion

This work presents and validates the ARCTIC (Adaptive Real-time Cryogenic Tank Inference and Correction) framework for accurate forecasting of thermodynamic behavior in cryogenic storage tanks. By integrating a nodal model with a real-time monitoring agent, ARCTIC derives data-driven correlations from real-time sensor data, effectively compensating for model imperfections in complex physical phenomena, such as rapid sloshing effects, while adapting to uncertainties in boundary conditions and unforeseen cryogenic liquid behaviors.

A key strength of ARCTIC lies in its non-intrusive design and generalized correction mechanism. Rather than tuning individual model parameters, such as interfacial heat transfer coefficients, the monitoring agent directly maps nodal model predictions to sensor measurements using lightweight, robust data-driven correlations. This strategy enables ARCTIC to inherently capture the cumulative effects of modeling inaccuracies, boundary fluctuations, and previously unmodeled phenomena. The modular design ensures compatibility with a wide range of nodal codes and accommodates future improvements in simulation fidelity.

Extensive synthetic testing demonstrates the framework's ability to maintain high forecasting accuracy across diverse thermodynamic scenarios, including sloshing, self-pressurization, and rapid transients. The framework efficiently handles combined effects of multiple unmodeled phenomena, boundary fluctuations, and sensor noise. Validation against multiple NASA experimental datasets, including long-duration self-pressurization experiments and rapid transients induced by sloshing, confirms that ARCTIC can deliver robust, highly accurate predictions under fast-changing, noisy, realistic conditions. A demonstration of the adaptive forecasting and control scenario further illustrates the framework's adaptability by handling a scenario in which the offline simulation fails to predict an over-pressurization event. ARCTIC successfully anticipates the deviation, enables the scheduling of a new control action, and re-synchronizes its predictions in real time, demonstrating its utility in practical autonomous control applications. Based on the above results, once the ARCTIC is integrated with the control system, the auto-calibration frequency could be adapted to different actions and conditions. For example, forecasting could be updated more frequently when a rapid event, such as maneuver-induced sloshing, is approaching. In contrast, the forecasting window could be extended during stable long-term storage periods. In this way, the approach could dynamically balance efficiency and accuracy according to real-time system requirements.

Future extensions of ARCTIC may focus on two complementary directions. First, sensitivity studies (see Appendix II) suggest that using multiple observation and forecasting windows simultaneously to generate slightly different predictions that reflect different assumptions about system dynamics and noise levels, providing a basis for real-time uncertainty quantification. Development of an auxiliary agent that can adaptively select the optimal observation and forecasting windows would also be highly valuable. Second, for complex scenarios where new control actions significantly alter system behavior, ARCTIC could be coupled with rapid re-simulation routines that update the offline model using data-informed initial conditions. Because full nodal re-simulations may be too slow for onboard use, neural-network-based surrogate models offer a promising path for enabling fast re-simulation within the ARCTIC framework.

In summary, ARCTIC provides an efficient and robust data-informed forecasting solution for real-time cryogenic storage tank thermodynamic state prediction. Its ability to adapt to unforeseen events, correct modeling gaps, and support autonomous decision-making makes it a promising tool for advancing cryogenic fluid management systems, particularly for deep-space missions where communication delays make autonomous and predictive control essential.

## Appendix I. Physical Significance of Data-Driven Correlations

This section discusses the physical significance of the data-driven correlations constructed using the algorithm described in Section 2, as well as the necessity of the auto-calibration and observation mechanisms in the ARCTIC framework for real-time data assimilation in control systems.

In the main text, three major challenges in cryogenic tank thermodynamic state prediction were discussed:

1. The complex space environment can trigger multiple dynamic events, such as pressurization, venting, and sloshing, each governed by distinct physical mechanisms that require different physical or empirical models for accurate representation.

2. Measurement noise, together with uncertainties in initial and boundary conditions, contributes to deviations between offline predictions and real-time measurement data.

3. Moreover, measurement data collected during one event (e.g., pressurization) is not useful to reliably correct the offline predictions corresponding to another event (e.g., sloshing) due to their fundamentally different physics.

To address these challenges, the ARCTIC framework is designed to continuously capture the current relationship between offline predictions and measurements, updating the mapping dynamically using only the most recently collected data that accurately represent the current system behavior.

Figure A1 illustrates the phase curves between offline predictions and measurement data for two representative cases: (a) the sequential synthetic scenario discussed in Section 4.3, and (b) the sloshing experiment from the K-Site facility presented in Section 5.3. In each plot, the x-axis represents the offline prediction $x_O(t)$ and the y-axis represents the corresponding measurement $x_M(t)$. The phase curves, obtained by pairing $\left(x_O(t), x_M(t)\right)$, visualize the instantaneous functional relationship required to map offline predictions into real-time measurements.

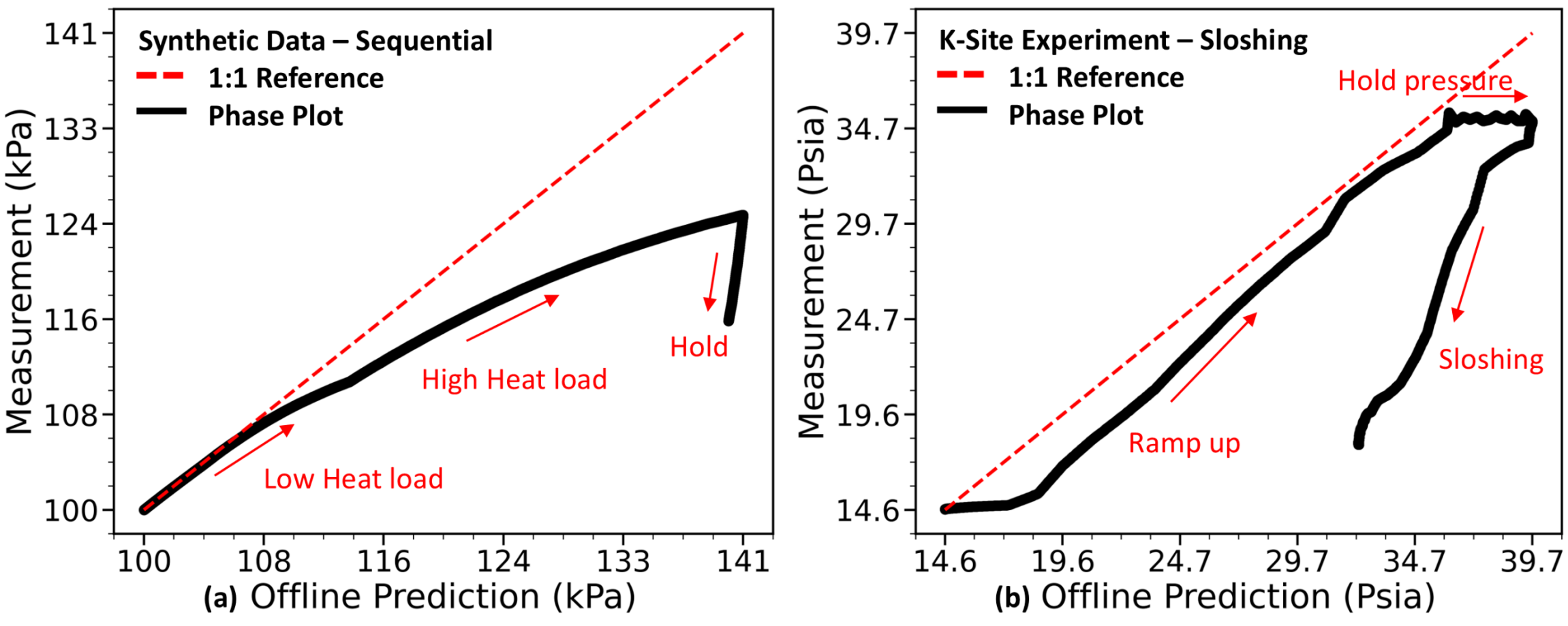

Figure A1. Phase curves of the offline predictions and measurement data. The x-axis represents the offline prediction $x_O(t)$ and the y-axis represents the measurement data $x_M(t)$. The curves are plotted by pairing $\left(x_O(t), x_M(t)\right)$. (a) Phase curve for the sequential synthetic problem described in Section 4.3, including three operating segments corresponding to low-heat-load pressurization, high-heat-load pressurization, and holding. (b) Phase curve for the K-Site sloshing experiment described in Section 5.3, exhibiting strongly nonlinear and noisy behavior due to regime transitions and measurement uncertainty.

In Figure A1(a), without the effect of the measurement noise, three distinct nonlinear relationships are observed between $x_O(t)$ and $x_M(t)$, corresponding to the low-heat-load pressurization, high-heat-load pressurization, and holding phases, each associated with a unique configuration of boundary conditions. In contrast, Figure A1(b) reveals a far more complex and noisy mapping between $x_O(t)$ and $x_M(t)$ for the K-Site sloshing experiment, reflecting the highly nonlinear and transient nature of the process.

These results highlight two key insights:

1. It is impractical to derive a universal data-driven correlation capable of correcting the offline predictions across all possible events.

2. Measurement data from one regime cannot be reused to correct predictions in another regime, because the mapping between $x_O(t)$ and $x_M(t)$ is no longer the same.

Therefore, for real-time data assimilation, the data-driven correlation must be updated frequently, and only the most recent data are reliable for correction. When an entirely new event occurs, an observation period must first be initiated to collect representative data before performing further corrections. In practice, it was found that a linear mapping provides the most stable and efficient performance. When employed within the ARCTIC framework, the linear correlation effectively linearizes the local segment of the phase curve. Through periodic auto-calibration, this local linearization is continuously updated, ensuring its validity as system conditions evolve.

**Appendix II. Sensitivity Study of Observation and Forecasting Windows**

Sensitivity studies are performed to examine the effects of different combinations of the observation window $\Delta t_D$ and the forecasting window $\Delta t_{AC}$ within the ARCTIC framework. In ARCTIC, observation periods introduce blackout intervals during which real-time forecasting is temporarily suspended while new measurement data are assimilated. Outside these observation periods, the data-informed forecast is guaranteed to remain the accuracy up to a predefined threshold; otherwise, a new observation is automatically triggered. Consequently, the objective of the sensitivity study is not to assess prediction accuracy, which is enforced by design, but rather to evaluate how to select pairs ($\Delta t_D$ and $\Delta t_{AC}$) of window combinations to minimize observation frequency and blackout duration while maintaining long-range forecasting capability.

Three representative problems are considered:

(1) a slow-varying case without measurement noise,

(2) a slow-varying case with high noise, and

(3) a fast-changing case with measurement noise.

These studies provide practical guidance for selecting appropriate window sizes under different operating conditions.

The first sensitivity study focuses on the long-term self-pressurization problem without measurement noise (Section 4.1). Figure A2 presents a comprehensive parameter scan in which each subplot shows pressure evolution as a function of time, while the forecasting windows $\Delta t_{AC}$ and observation windows $\Delta t_D$ are varied parametrically across columns and rows, respectively. Specifically, columns correspond to increasing forecasting windows $\Delta t_{AC}$ and rows corresponding to increasing observations windows $\Delta t_D$.

Although the data-informed predictions outside the observation periods (gray shaded regions) appear nearly identical across parameter combinations, important differences emerge in the frequency and duration of blackout intervals. For example, configurations with larger observation windows (e.g., Figure A2(b)) trigger observations more frequently, resulting in repeated forecasting interruptions. This behavior is undesirable from an operational perspective, even though short-term prediction accuracy remains high. This trend arises because, in the absence of measurement noise, short observation windows more effectively capture the locally linearized mapping between offline predictions and measurements. In contrast, longer observation windows average data over a broader temporal span, during which nonlinear effects accumulate and cause the correction to exceed the admissible error threshold more frequently. As a result, additional observations are triggered, increasing blackout frequency.

Overall, the results indicate that for slow-varying, noise-free problems, the observation window $\Delta t_D$ can be selected to be very short, while the forecasting window $\Delta t_{AC}$ can be extended significantly without compromising prediction accuracy or forecasting continuity.

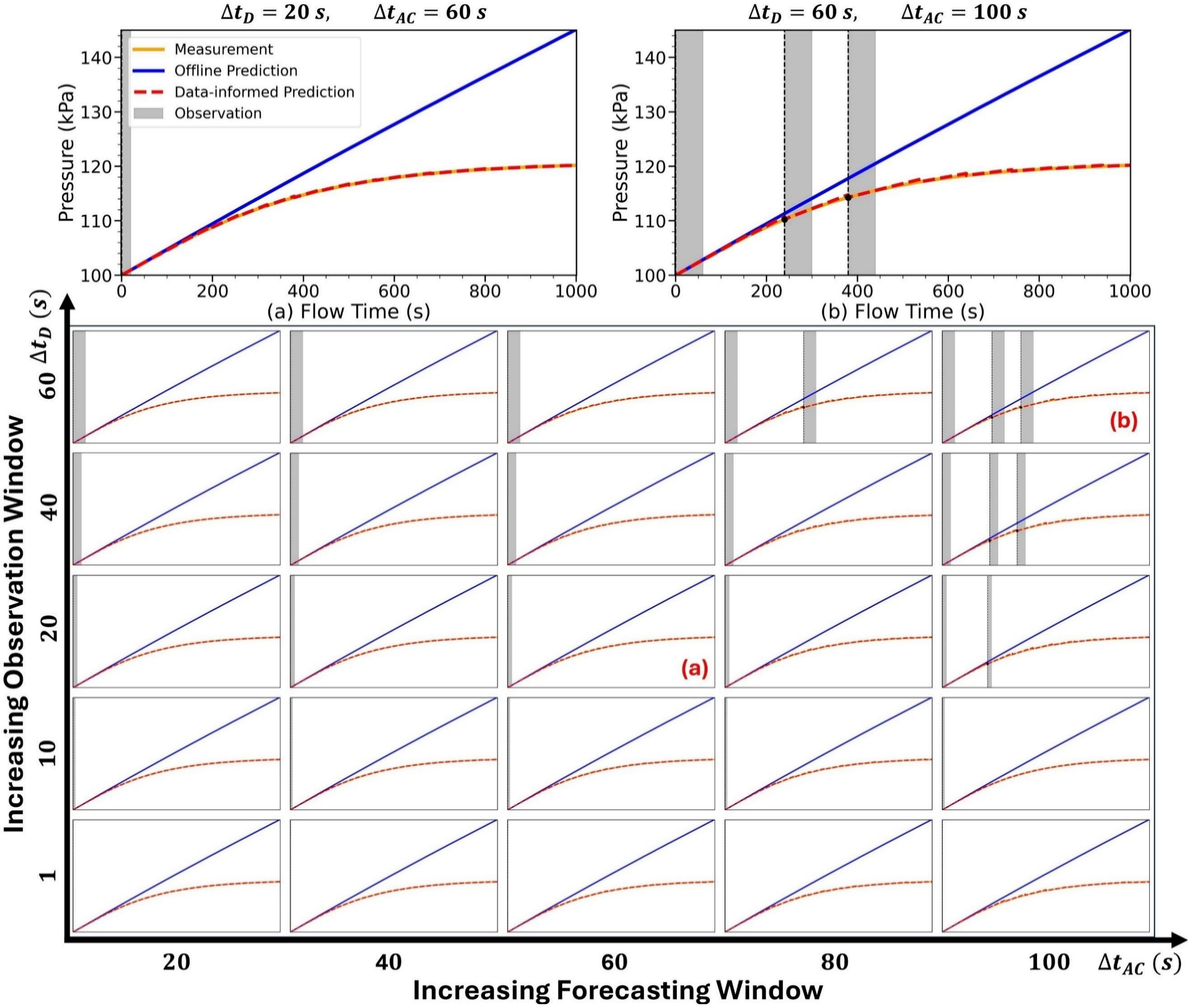

Figure A2. Sensitivity study of observation windows $\Delta t_D$ and forecasting windows $\Delta t_{AC}$ for the long-term self-pressurization problem without measurement noise. Each subplot shows pressure (y-axis) as a function of time (x-axis). Gray shaded regions indicate observation periods, during which ARCTIC temporarily

suspends real-time forecasting (blackout intervals). The top row shows two zoomed-in configurations: (a) $\Delta t_D = 20s, \Delta t_{AC} = 60s$ , (b) $\Delta t_D = 60s, \Delta t_{AC} = 100s$ , illustrating how window selection affects observation triggering behavior. The lower panel presents a comprehensive parameter scan. Columns correspond to increasing forecasting windows $\Delta t_{AC} = \{20,40,60,80,100\}s$, while rows correspond to increasing observation windows $\Delta t_D = \{1,10,20,40,60\}s$. All subplots share identical axis ranges to enable direct visual comparison; axis labels are omitted for compactness. While data-informed predictions outside the observation periods closely follow the measurements across all cases, larger observation windows lead to more frequent observation triggering and increased blackout frequency.

The sensitivity behavior for the same long-term self-pressurization problem, but with high measurement noise is presented in Figure A3. Above the parameter scan panel, two representative parameter combinations are zoomed in. In Figure A3(a), a very short observation window fails because measurement noise dominates the signal, causing the fitted data-driven correction to become unstable and unrepresentative of the underlying system dynamics. In contrast, longer observation windows (Figure A3(b)) include sufficient data to average out noise, resulting in more stable forecasting behavior. Compared with the noise-free case shown in Figure A2, the nonlinear effects associated with long observation windows become secondary relative to noise amplitude. As a result, longer observation windows are preferable in noisy environments, even though they may increase observation duration.

The lower panel of Figure A3 shows a comprehensive parameter scan in which columns correspond to increasing forecasting windows $\Delta t_{AC}$, and rows correspond to increasing observation windows $\Delta t_D$. The results confirm that, for slow-varying systems with high measurement noise, longer observation windows are required to ensure stable and representative data-driven corrections, whereas the forecasting window can still be extended without significant degradation in performance.

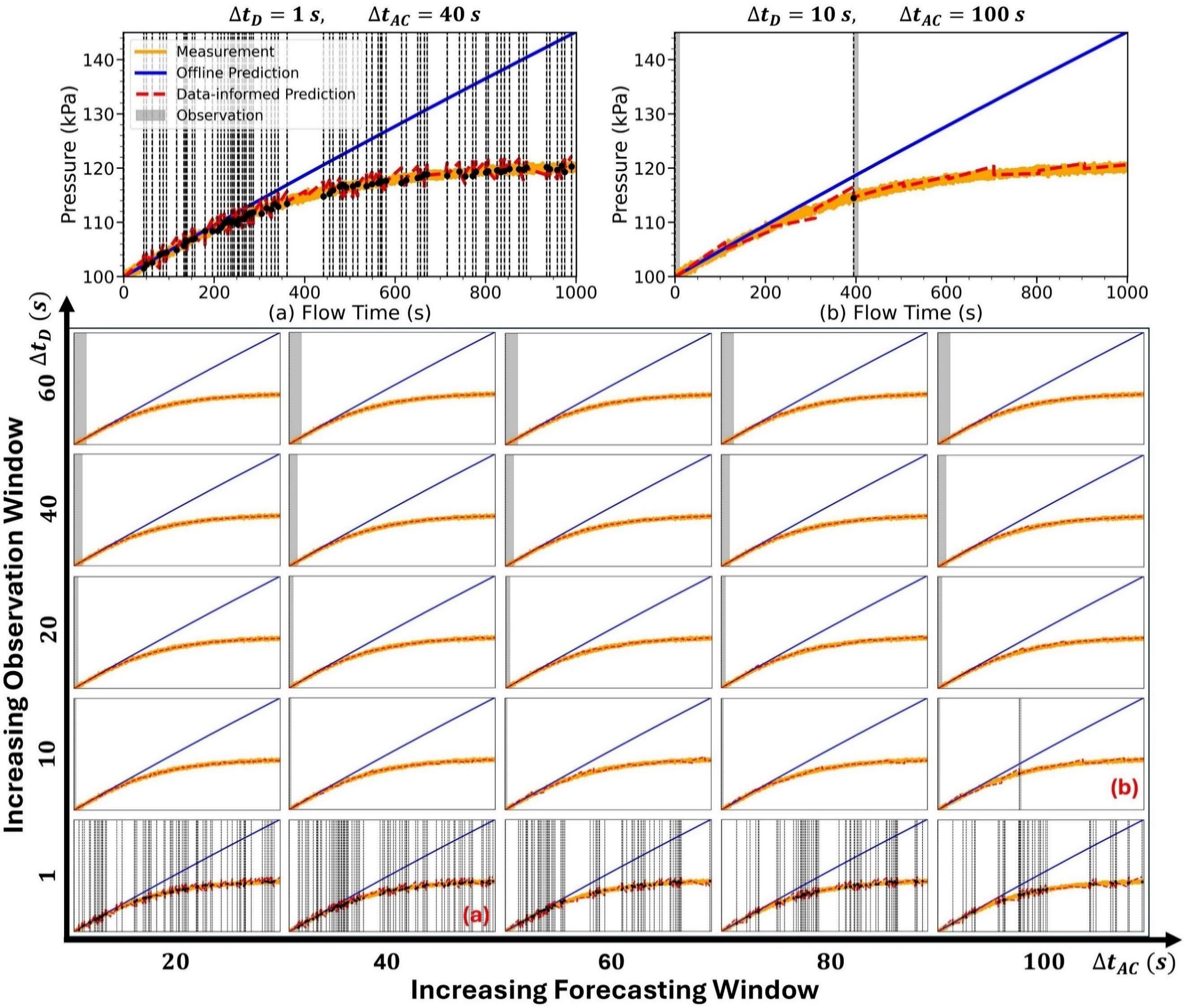

Figure A3. Sensitivity study of observation windows $\Delta t_D$ and forecasting windows $\Delta t_{AC}$ for the long-term self-pressurization problem with high measurement noise. Each subplot shows pressure (y-axis) as a function of time (x-axis). Gray shaded regions indicate observation periods, during which ARCTIC temporarily suspends real-time forecasting (blackout intervals). The top row shows two zoomed-in configurations: (a) $\Delta t_D = 1s, \Delta t_{AC} = 40s$, (b) $\Delta t_D = 10s, \Delta t_{AC} = 100s$, illustrating how window selection affects observation triggering behavior. The lower panel presents a comprehensive parameter scan. Columns correspond to increasing forecasting windows $\Delta t_{AC} = \{20,40,60,80,100\}s$, while rows correspond to increasing observation windows $\Delta t_D = \{1,10,20,40,60\}s$. All subplots share identical axis ranges to enable direct visual comparison; axis labels are omitted for compactness.

The sensitivity results for the K-Site sloshing experiment described in Section 5.3 are presented in Figure A4. This case involves multiple time scales, such as the slow evolution during pressure holding and the rapid sloshing-induced transient. The results reflect this multi-scale behavior. Long observation windows mix data from different physical regimes and are heavily influenced by nonlinear and fast-changing sloshing behavior. At the same time, sufficiently frequent auto-calibration (short forecasting windows) is necessary to capture the rapid transient. However, when the updating is too frequent and the observation windows are short (e.g., the $(\Delta t_{AC} = 2, \Delta t_D = 0.5)$ combination), measurement noise and multi-scale coupling can degrade the performance. The best performance is obtained using the $(\Delta t_{AC} = 4, \Delta t_D = 0.5)$ combination, which uses a short observation window to capture the instantaneous mapping while providing a forecasting horizon long enough to avoid excessive observation periods.

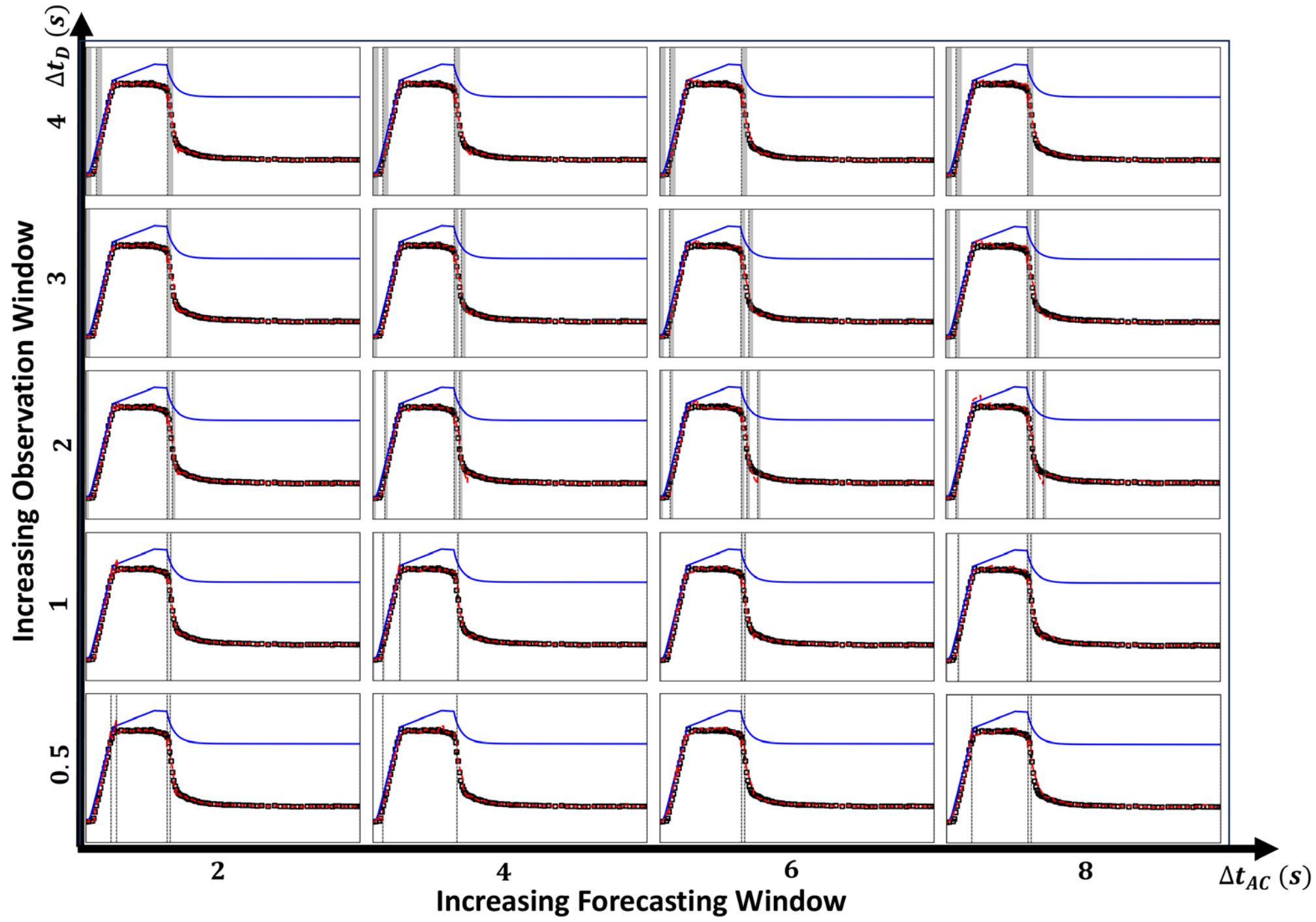

Figure A4. Sensitivity study of the forecasting windows $\Delta t_{AC}$ and observation windows $\Delta t_D$ for the K-Site sloshing experiment. Different columns correspond to increasing forecasting windows $\Delta t_{AC}$ (2s, 4s, 6s, and 8s), while different rows correspond to increasing observation windows $\Delta t_D$ (0.5s, 1s, 2s, 3s, and 4s). All subplots share the same axis and axis ranges as Figure 12, and axis labels are omitted here to facilitate direct visual comparison across subplots. The results demonstrate how new observations are triggered due to the different combinations of parameters.

To summarize, the sensitivity studies demonstrate that in ARCTIC the observation window should be chosen as short as possible to capture the local linearized mapping, but not so short that it becomes dominated by measurement noise. The forecasting window should be chosen based on the expected event type and the reaction time required by the control system: shorter forecasting windows are suitable for fast-changing events such as sloshing, while longer forecasting windows are appropriate for slowly evolving processes such as pressurization. These insights provide practical guidance for configuring ARCTIC in real-time applications and ensuring reliable performance under diverse operating conditions.

These sensitivity studies also suggest an interesting direction for future development. If ARCTIC were to employ multiple observation and forecasting windows simultaneously, each combination would produce slightly different predictions. These predictions could be interpreted as representing different assumptions about system behaviors, such as slow or fast system changes, low or high measurement noise. Combining these predictions may provide not only a more comprehensive description of future thermodynamic states, but also a path toward uncertainty quantification for real-time forecasting. This capability would further enhance ARCTIC as a reliable engineering-level prediction tool for cryogenic tanks.

**Appendix III. Pseudo Code for the Fitting Algorithm in ARCTIC**

This appendix summarizes the polynomial regression procedure used in ARCTIC to construct the data-driven correlation between the offline predictions and the measurement data. The algorithm is expressed here in a simplified pseudo-code form for clarity and is implemented in ARCTIC using SciPy library [42].

**Algorithm: Polynomial Regression**

// Measurement data over the observation window $X_M = [x_{M,1}, x_{M,2}, \dots, x_{M,N}]^\top$

// Offline prediction data over the same window $X_O = [x_{O,1}, x_{O,2}, \dots, x_{O,N}]^\top$

// Polynomial order $p$, Regression coefficients $\theta = [\theta_0, \theta_1, \theta_2 \dots, \theta_p]^\top$

**1** Data-Driven Correlation $g_\theta(x; \theta)$:

  // Construct polynomial feature of order p

**2**  **For** $i = 1,2, \dots, N$ **do**

**3**   $\Phi(i) = [1, x(i), x(i)^2, \dots, x(i)^p]$

  // Solve the regression coefficients $\theta^*$ by least-squares

**5**  $\theta^* = \arg\min_{\theta} \|\Phi\theta - X_M\|^2$

  // Return the mapping function

**6**  **Return** $g_{\theta^*}(x; \theta^*)$